\documentclass[aps,prd,groupedaddress,nofootinbib,amssymb,eqsecnum,epsfig]{revtex4}
\usepackage{graphicx}
\usepackage{bm}
\usepackage{amsmath}
\usepackage{color}
\usepackage{amsfonts}

\begin{document}
\newcommand{\newc}{\newcommand}

\newcommand{\ben}{\begin{eqnarray}}
\newcommand{\een}{\end{eqnarray}}
\newc{\be}{\begin{equation}}
\newc{\ee}{\end{equation}}
\newc{\ba}{\begin{eqnarray}}
\newc{\ea}{\end{eqnarray}}
\newc{\bea}{\begin{eqnarray*}}
\newc{\eea}{\end{eqnarray*}}
\newc{\D}{\partial}
\newc{\ie}{{\it i.e.} }
\newc{\eg}{{\it e.g.} }
\newc{\etc}{{\it etc.} }
\newc{\etal}{{\it et al.}}
\newcommand{\nn}{\nonumber}
\newc{\ra}{\rightarrow}
\newc{\lra}{\leftrightarrow}
\newc{\lsim}{\buildrel{<}\over{\sim}}
\newc{\gsim}{\buildrel{>}\over{\sim}}

\title{Anisotropic cosmological solutions 
in massive vector theories}

\author{
Lavinia Heisenberg$^{1}$, 
Ryotaro Kase$^{2}$, and 
Shinji Tsujikawa$^{2}$}

\affiliation{
$^1$Institute for Theoretical Studies, ETH Zurich, Clausiusstrasse 47, 8092 Zurich, Switzerland\\
$^2$Department of Physics, Faculty of Science, Tokyo University of Science, 1-3, Kagurazaka,
Shinjuku-ku, Tokyo 162-8601, Japan}

\date{\today}

\begin{abstract}

In beyond-generalized Proca theories including the 
extension to theories higher than second order, 
we study the role of a spatial component $v$ of 
a massive vector field on the anisotropic 
cosmological background. 
We show that, as in the case of the isotropic 
cosmological background, there is no additional 
ghostly degrees of freedom associated with 
the Ostrogradski instability. 
In second-order generalized Proca theories we find the 
existence of anisotropic solutions on which the ratio 
between the anisotropic expansion rate $\Sigma$ 
and the isotropic expansion rate $H$ remains 
nearly constant in the radiation-dominated epoch.
In the regime where $\Sigma/H$ is constant, 
the spatial vector component $v$ works as 
a dark radiation with the equation of state 
close to $1/3$. 
During the matter era, the ratio $\Sigma/H$ 
decreases with the decrease of $v$.
As long as the conditions $|\Sigma| \ll H$ and 
$v^2 \ll \phi^2$ are satisfied around the onset of 
late-time cosmic acceleration, 
where $\phi$ is the temporal vector component, 
we find that the solutions approach the isotropic 
de Sitter fixed point ($\Sigma=0=v$)
in accordance with the cosmic no-hair conjecture. 
In the presence of $v$ and $\Sigma$ the early 
evolution of the dark energy equation of state 
$w_{\rm DE}$ in the radiation era is different from 
that in the isotropic case, but the approach to 
the isotropic value $w_{\rm DE}^{{\rm (iso)}}$ typically occurs
at redshifts $z$ much larger than 1.
Thus, apart from the existence of dark radiation, the 
anisotropic cosmological dynamics at low redshifts is similar to 
that in isotropic generalized Proca theories.
In beyond-generalized Proca theories 
the only consistent solution to avoid the divergence 
of a determinant of the dynamical system corresponds to 
$v=0$, so $\Sigma$ always decreases in time.

\end{abstract}


\maketitle

\section{Introduction}

Cosmology is facing a challenge of revealing the origin of 
unknown dark components dominating the present Universe.
The standard cosmological model introduces a pure cosmological constant 
$\Lambda$ to the field equations of General Relativity (GR), 
and additionally a non-baryonic dark matter component in the form of 
non-relativistic particles, known as cold dark matter.
This rather simple model is overall consistent with the tests at cosmological scales by the Cosmic Microwave Background (CMB) 
anisotropies \cite{CMB}, by the observed matter distribution 
in large-scale structures \cite{LSS}, and by the type Ia 
supernovae \cite{SNIa}.

In the prevailing view the cosmological constant should arise from 
the vacuum energy density, which can be estimated by 
using techniques of the standard quantum field theory.
Lamentably, the theoretically estimated value of vacuum energy 
is vastly larger than the observed dark energy scale.
This is known as the cosmological constant problem \cite{Weinberg}. 
In this context, infrared modifications of gravity have been widely studied in the hope to screen the cosmological constant via a high-pass filter, naturally arising in higher dimensional set-ups, in massive gravity \cite{mgravity} 
or non-local theories \cite{nonlocal}. 

On the same footing as tackling the cosmological constant problem, infrared modifications of gravity may yield accelerated expansion of the Universe due to the presence of new physical degrees of freedom, providing an alternative framework for dark energy \cite{review}. The simplest realization is in form of an additional scalar degree of freedom, that couples minimally 
to gravity \cite{quin}. 
Richer phenomenology can be achieved by allowing for self-interactions of the scalar field or non-minimal couplings to gravity \cite{stensor}. 
Extensively studied classes are the Galileon \cite{Galileon1,Galileon2} and Horndeski \cite{Horndeski1} theories, whose constructions rely on the condition 
of keeping the equations of motion up to second order \cite{Horn2}.
Easing this restriction allows for interactions with higher-order equations of motion, but it is still possible to construct theories that propagate
only one scalar degree of freedom (DOF)--Gleyzes-Langlois-Piazza-Vernizzi 
(GLPV) theories \cite{GLPV} (see also Refs.~\cite{Hami} for the discussion 
of the propagating DOF).
These beyond-Horndeski interactions generalize the previous ones 
and offer richer phenomenology \cite{GLPVcosmo,GLPVsph}.

Even if the extension in the form of an additional scalar field is the simplest and most explored one, the inclusion of additional vector fields has been taken into consideration as well \cite{Barrow,Jimenez13,TKK,Fleury,Hull}. 
The Maxwell field with the standard kinetic term is a gauge-invariant vector field with two propagating transverse modes. 
The attempt to construct derivative self-interactions for a massless, 
Lorentz-invariant vector field yielded a no-go result \cite{Mukoh}, but this 
can be overcome by breaking the gauge invariance. The simplest way of breaking gauge invariance is to include a mass term for the vector field, which is known as the Proca field. The inclusion of allowed derivative self-interactions and non-minimal couplings to gravity results in the generalized Proca 
theories with second-order equations of 
motion \cite{Heisenberg,Tasinato,Allys,Jimenez2016}.
As in the original Proca theory, there are still the three physical DOFs, one of them being the longitudinal mode and the other two corresponding to the standard transverse modes (besides two tensor polarizations). The phenomenology of (sub classes of) 
generalized Proca theories has been extensively studied in 
Refs.~\cite{scvector,Chagoya,cosmo,Geff}.

As in the GLPV extension of scalar Horndeski theories, 
relaxing the condition of second-order equations of motions in 
generalized Proca theories allows us to construct new higher-order 
vector-tensor interactions \cite{HKT} without the Ostrogradski 
instability \cite{Ostro}. In Ref.~\cite{HKT} 
it was shown that, even in the presence of interactions beyond 
the domain of generalized Proca theories, there are no additional 
DOFs associated with the  Ostrogradski ghost on the isotropic 
Friedmann-Lema\^{i}tre-Robertson-Walker (FLRW) background.
In fact, the Hamiltonian ${\cal H}$ is equivalent to 0 due to the existence of 
a constraint that removes the appearance of a ghostly DOF.

In Refs.~\cite{cosmo,Geff} the cosmology of generalized Proca 
theories was studied by introducing the temporal component $\phi(t)$ 
of a vector field $A^{\mu}$ at the background level (where 
$t$ is the cosmic time). 
There the spatial vector component was treated as a perturbation 
on the FLRW background. 
In concrete dark energy models it was found that $\phi(t)$ grows
toward a de Sitter attractor, whereas the vector perturbation decays 
after entering the vector sound horizon.
Thus the analysis of Refs.~\cite{cosmo,Geff} is self-consistent, but 
it remains to see whether or not the spatial vector component $v(t)$ as large as 
$\phi(t)$ can modify the cosmological dynamics. 
If $v(t)$ is non-negligible relative to $\phi(t)$, we need to take into 
account the spatial shear $\sigma(t)$ in the metric as well. 
In the context of anisotropic inflation, for example, it is known that 
there are cases in which the ratio between the anisotropic and isotropic 
expansion rates stays nearly constant \cite{aniinf}. 
In generalized Proca theories, we would like to clarify the behavior of $v(t)$ and the anisotropic expansion rate 
$\Sigma(t)=\dot{\sigma}(t)$ during the cosmic 
expansion history from 
the radiation era to the late-time accelerated epoch.

In beyond-generalized Proca theories, it is important to study
whether the Ostrogradski ghost appears or not on the anisotropic cosmological background. In this paper, we show the absence of additional ghostly DOF by explicitly 
computing the 
Hamiltonian in the presence of $v$ and $\Sigma$.
An interesting property of beyond-generalized Proca theories 
with quartic and quintic Lagrangians is that the only consistent 
solution free from a determinant singularity of the dynamical 
system corresponds to $v=0$. 
In this case, the cosmological dynamics can be well described 
by the isotropic one studied in Ref.~\cite{HKT}.

We organize our paper as follows. 
In Sec.~\ref{eqsec} we derive the Hamiltonian and the full equations 
of motion in beyond-generalized Proca theories on the anisotropic 
cosmological background.
In Sec.~\ref{anasec} we analytically estimate the evolution of 
$v$ and $\Sigma$ in the radiation/matter eras and in the 
de Sitter epoch. 
In Sec.~\ref{numesec} we perform the numerical study to 
clarify the cosmological dynamics for both generalized and 
beyond-generalized Proca theories in detail, paying particular 
attention to the evolution of the dark energy equation of state 
$w_{\rm DE}$ as well as $v$ and $\Sigma$.
Sec.~\ref{consec} is devoted to conclusions.

\section{Equations of motion on the anisotropic 
cosmological background}
\label{eqsec}

The beyond-generalized Proca theories \cite{HKT}, which encompass 
the second-order generalized Proca theories as a 
specific case \cite{Heisenberg,Jimenez2016}, 
are described by the action 
\be
S=\int d^4x \sqrt{-g} \left( {\cal L}_F
+{\cal L}_M 
+ \sum_{i=2}^{6} {\cal L}_i
+{\cal L}^{\rm N}\right)\,, 
\label{action}
\ee
where $g$ is the determinant of metric $g_{\mu \nu}$, 
${\cal L}_F$ is the standard Maxwell term given by 
\be
{\cal L}_F=-\frac14 F_{\mu \nu} F^{\mu \nu}\,,
\label{LF}
\ee
with $F_{\mu \nu}=
\nabla_{\mu}A_{\nu}-\nabla_{\nu}A_{\mu}$, 
and ${\cal L}_M $ is the matter Lagrangian 
density ($\nabla_{\mu}$ is the covariant 
derivative operator).

The second-order generalized Proca theories, 
which break the U(1) gauge-invariance 
in the presence of a vector mass term $m$, 
correspond to the Lagrangian densities 
${\cal L}_{2,3,4,5,6}$ in Eq.~(\ref{action}). 
They are given, respectively, by \cite{Heisenberg,Jimenez2016}
\ba
{\cal L}_2 &=& G_2(X)\,, 
\label{L2}\\
{\cal L}_3 &=& G_3(X) \nabla_{\mu}A^{\mu}\,,
\label{L3}\\
{\cal L}_4 &=& 
G_4(X)R+
G_{4,X}(X) \left[ (\nabla_{\mu} A^{\mu})^2
-\nabla_{\rho}A_{\sigma}
\nabla^{\sigma}A^{\rho} \right]
+\frac12 g_4(X) F_{\mu \nu} F^{\mu \nu}\,,\label{L4} \\
{\cal L}_5 &=& 
G_{5}(X) G_{\mu \nu} \nabla^{\mu} A^{\nu}
-\frac16 G_{5,X}(X) [ (\nabla_{\mu} A^{\mu})^3
-3\nabla_{\mu} A^{\mu}
\nabla_{\rho}A_{\sigma} \nabla^{\sigma}A^{\rho} 
+2\nabla_{\rho}A_{\sigma} \nabla^{\gamma}
A^{\rho} \nabla^{\sigma}A_{\gamma}] \nonumber \\
& &-g_5(X) \tilde{F}^{\alpha \mu}
{\tilde{F^{\beta}}}_{\mu} \nabla_{\alpha} A_{\beta}\,,
\label{L5}\\
{\cal L}_6 &=& G_6(X) L^{\mu \nu \alpha \beta} 
\nabla_{\mu}A_{\nu} \nabla_{\alpha}A_{\beta}
+\frac12 G_{6,X}(X) \tilde{F}^{\alpha \beta} \tilde{F}^{\mu \nu} 
\nabla_{\alpha}A_{\mu} \nabla_{\beta}A_{\nu}\,,
\label{L6}
\ea
where $G_{2,3,4,5,6}$ and $g_{4,5}$ are functions 
of $X=-A_{\mu} A^{\mu}/2$ with the notation 
of partial derivatives as 
$G_{i,X} \equiv \partial G_i/\partial X$, 
$R$ is the Ricci scalar, and $G_{\mu \nu}$ 
is the Einstein tensor. 
The original massive Proca Lagrangian corresponds to 
$G_2(X)=m^2X$.
The quantities $L^{\mu \nu \alpha \beta}$ and 
$\tilde{F}^{\mu \nu}$ are the double dual Riemann 
tensor and the dual strength tensor
defined, respectively, by
\be
L^{\mu \nu \alpha \beta}=\frac14 {\cal E}^{\mu \nu \rho \sigma} 
{\cal E}^{\alpha \beta \gamma \delta} R_{\rho \sigma \gamma \delta}\,,
\qquad
\tilde{F}^{\mu \nu}=\frac12 {\cal E}^{\mu \nu \alpha \beta}
F_{\alpha \beta}\,,
\ee
where ${\cal E}^{\mu \nu \rho \sigma}$ is the 
Levi-Civita tensor and $R_{\rho \delta \gamma \delta}$ 
is the Riemann tensor. 
The Maxwell term (\ref{LF}) and the last term of 
Eq.~(\ref{L4}) can be absorbed into the Lagrangian 
density ${\cal L}_2$ by allowing the dependence of 
${\cal L}_F=-F_{\mu \nu}F^{\mu \nu}/4$ as 
$G_2=G_2(X,{\cal L}_F)$ \cite{Heisenberg,Allys,Geff}.
We separate the ${\cal L}_F$ dependence from ${\cal L}_2$ 
because this allows us to see the kinetic term  
of a spatial component $v$ of the vector field explicitly 
in the Lagrangian.
It is also possible to include the dependence of the term 
$Y=A^{\mu}A^{\nu} {F_{\mu}}^{\alpha} 
F_{\nu \alpha}$ in ${\cal L}_2$ \cite{Geff}.
On the anisotropic cosmological background studied  
in this paper, the quantity $Y$ can be expressed 
in terms of $X$ and ${\cal L}_F$ as
$Y=4X{\cal L}_F$, so we do not take into account such dependence.

The Lagrangian density ${\cal L}^{\rm N}$ in Eq.~(\ref{action}) 
corresponds to the one beyond the domain of second-order 
generalized Proca theories \cite{HKT}. 
This is given by the sum of four contributions
\be
{\cal L}^{\rm N}=
{\cal L}_4^{\rm N}+{\cal L}_5^{\rm N}+
\tilde{{\cal L}_5^{\rm N}}+
{\cal L}_6^{\rm N}\,,
\ee
where 
\ba
\hspace{-1.2cm}
& &{\cal L}_4^{\rm N}
=f_4 (X)\hat{\delta}_{\alpha_1 \alpha_2 \alpha_3 \gamma_4}^{\beta_1 \beta_2\beta_3\gamma_4}
A^{\alpha_1}A_{\beta_1}
\nabla^{\alpha_2}A_{\beta_2} 
\nabla^{\alpha_3}A_{\beta_3}\,, \label{L4N}\\
\hspace{-1.2cm}
& &{\cal L}_5^{\rm N}
=
f_5 (X)\hat{\delta}_{\alpha_1 \alpha_2 \alpha_3 \alpha_4}^{\beta_1 \beta_2\beta_3\beta_4}
A^{\alpha_1}A_{\beta_1} \nabla^{\alpha_2} 
A_{\beta_2} \nabla^{\alpha_3} A_{\beta_3}
\nabla^{\alpha_4} A_{\beta_4}\,,\label{L5N} \\
\hspace{-1.2cm}
& &\tilde{{\cal L}}_5^{\rm N}
=
\tilde{f}_{5} (X)
\hat{\delta}_{\alpha_1 \alpha_2 \alpha_3 \alpha_4}^{\beta_1 \beta_2\beta_3\beta_4}
A^{\alpha_1}A_{\beta_1} \nabla^{\alpha_2} 
A^{\alpha_3} \nabla_{\beta_2} A_{\beta_3}
\nabla^{\alpha_4} A_{\beta_4}\,,
\label{L5Nd} \\
\hspace{-1.2cm}
& &
{\cal L}_6^{\rm N}
=f_{6}(X)
 \hat{\delta}_{\alpha_1 \alpha_2 \alpha_3 \alpha_4}^{\beta_1 \beta_2\beta_3\beta_4}
\nabla_{\beta_1} A_{\beta_2} \nabla^{\alpha_1}A^{\alpha_2}
\nabla_{\beta_3} A^{\alpha_3} \nabla_{\beta_4} A^{\alpha_4}\,,
\label{L6N}
\ea
with $\hat{\delta}_{\alpha_1 \alpha_2\gamma_3\gamma_4}^{\beta_1 \beta_2\gamma_3\gamma_4}={\cal E}_{\alpha_1 \alpha_2\gamma_3\gamma_4}
{\cal E}^{\beta_1 \beta_2\gamma_3\gamma_4}$, and 
the functions $f_{4,5,6}, \tilde{f}_{5}$ depend on $X$. 
Taking the limit $A^{\mu} \to \nabla^{\mu} \pi$, 
the Lagrangian densities ${\cal L}_4^{\rm N}$ and 
${\cal L}_5^{\rm N}$ of the scalar field $\pi$ 
are equivalent to those appearing in GLPV theories \cite{GLPV}.
In GLPV theories, such terms do not give rise to an extra
DOF associated with the Ostrogradski ghost. 
For the vector-field Lagrangian densities (\ref{L4N})-(\ref{L6N}) 
it was shown in Ref.~\cite{HKT} that additional propagating DOFs
to those appearing in second-order generalized Proca theories 
(one longitudinal mode and two transverse polarizations)
do not arise on the maximally symmetric space-time and for linear 
cosmological perturbations on the flat FLRW background.

In Refs.~\cite{cosmo,Geff} the cosmology in generalized Proca theories was studied 
on the flat FLRW background under the assumption that the vector field 
$A^{\mu}$ has a time-dependent temporal component $\phi(t)$ alone.
The spatial part of $A^{\mu}$ was treated as the perturbations on the 
FLRW background. In this paper, we would like to explicitly include the 
spatial component $v(t)$ of $A^{\mu}$ besides the temporal 
component $\phi(t)$ already present at the background level. For concreteness we consider the vector field 
$A^i$ pointing to the $x$-direction. 
Since there is the rotational symmetry in the $(y, z)$ 
plane, we take the line-element in the following form \cite{aniinf}
\be
ds^2=-N^2(t)dt^2+e^{2\alpha(t)} \left[ e^{-4\sigma(t)}dx^2
+e^{2\sigma(t)} \left( dy^2+dz^2 \right) \right]\,,
\label{metric}
\ee
where $N(t)$ is the lapse, $e^{\alpha} \equiv a$ is the isotropic 
scale factor, and $\sigma$ characterizes the 
deviation from isotropy.
We write the vector field in the form 
\be
A^{\mu}=\left( \frac{\phi(t)}{N(t)}, 
e^{-\alpha(t)+2\sigma(t)}\,v(t), 0, 0 \right)\,,
\ee
in which case the term $X$ is given by 
\be
X=\frac12 \phi^2(t)-\frac12 v^2(t)\,.
\ee

Expanding the action (\ref{action}) for the line element (\ref{metric}) 
and integrating the second derivatives $\ddot{\alpha}$ and $\ddot{\sigma}$ 
by parts, we obtain the action $S=\int d^4 x\,L$ with the Lagrangian
\be
L=\frac{e^{3\alpha}}{N^3} \left( {\cal F}_1+{\cal F}_2\dot{\sigma}^2
+{\cal F}_3 \dot{v}^2+{\cal F}_4 \dot{v}v+{\cal F}_5v^2
+N^4{\cal L}_M \right)\,,
\label{Lag}
\ee
where a dot represents the derivative with 
respect to $t$, and
\ba
\hspace{-1.0cm}
{\cal F}_1 &=& 
N \left[ N^3 G_2+N^2 G_3(\dot{\phi}+3\dot{\alpha}\phi)
+6N\dot{\alpha}^2 (G_{4,X}\phi^2-G_4)
-G_{5,X}\dot{\alpha}^3 \phi^3
+6\dot{\alpha}^2 \phi^4 (Nf_4
+f_5\dot{\alpha}\phi) \right],\\
\hspace{-1.0cm}
{\cal F}_2 &=& 
N \left[ 6NG_4-6N G_{4,X}\phi^2+G_{5,X}\phi^3 
(3\dot{\alpha}+2\dot{\sigma})-6Nf_4\phi^4-6f_5 \phi^5 
(3\dot{\alpha}+2\dot{\sigma}) \right]
\,,\\
\hspace{-1.0cm}
{\cal F}_3 &=& \frac12 (1-2g_4)N^2 
-(\dot{\alpha}+\dot{\sigma}) \left[ 2Ng_5 \phi
-(\dot{\alpha}+\dot{\sigma}) 
\{ G_6+(G_{6,X}+2f_6)\phi^2 \} 
\right]
\,,\\
\hspace{-1.0cm}
{\cal F}_4 &=& 
\dot{\alpha} \left[ 2(G_{6,X}+2f_6)
\dot{\alpha}^2 \phi^2+N^2 (1+4G_{4,X}
+4f_4\phi^2-2g_4)-N\dot{\alpha}\phi 
(G_{5,X}-6f_5\phi^2+4g_5) 
\right] \nonumber \\
\hspace{-1.0cm}
&&+2G_6 (\dot{\alpha}-2\dot{\sigma})
(\dot{\alpha}+\dot{\sigma})^2
-4(G_{6,X}+2f_6)\phi^2\dot{\sigma}^3
+\phi [8Ng_5-NG_{5,X}-6(G_{6,X}+2f_6)
\dot{\alpha} \phi+6Nf_5\phi^2]\dot{\sigma}^2
\nonumber \\
\hspace{-1.0cm}
&&+2N[N(2G_{4,X}-1+2f_4\phi^2+2g_4)-\dot{\alpha} 
\phi (G_{5,X}-6f_5\phi^2-2g_5)]\dot{\sigma}\,,\\
\hspace{-1.0cm}
{\cal F}_5 &=& 
G_6(\dot{\alpha}-2\dot{\sigma})^2(\dot{\alpha}+\dot{\sigma})^2
+4(G_{6,X}+2f_6)\phi^2 \dot{\sigma}^4
-4\phi[2Ng_5-\phi(G_{6,X}\dot{\alpha}
+3Nf_5\phi+2f_6\dot{\alpha})]\dot{\sigma}^3 \nonumber \\
\hspace{-1.0cm}
& &
-[3(G_{6,X}+2f_6)\dot{\alpha}^2 \phi^2-2N^2 (1+3f_4\phi^2-2g_4)
+6Nf_5\phi^2 (\dot{\phi}-3\dot{\alpha}\phi)]\dot{\sigma}^2 \nonumber \\
\hspace{-1.0cm}
& & -2 [N\dot{\alpha}(N+6f_5\phi^2 \dot{\phi}-2Ng_4)
+(G_{6,X}+2f_6)\dot{\alpha}^3\phi^2
+2N^2 f_4 \phi \dot{\phi}-3Ng_5\dot{\alpha}^2 \phi]\dot{\sigma} \nonumber \\
\hspace{-1.0cm}
& &+\frac12 \dot{\alpha} [N\dot{\alpha} 
\{(1-2g_4)N-12\phi^2 (Nf_4+f_5\dot{\phi})\}
+2(G_{6,X}+2f_6)\dot{\alpha}^3 \phi^2
-8N^2 f_4\phi \dot{\phi}
-4N\dot{\alpha}^2 \phi(3f_5\phi^2+g_5)]\,.
\ea
In the isotropic case we have 
that $\sigma=0$ and $v=0$, 
so the Lagrangian (\ref{Lag}) reduces to 
$L=e^{3\alpha}({\cal F}_1+N^4 {\cal L}_M)/N^3$.
When the spatial anisotropy is present, the terms containing 
${\cal F}_{2,3,4,5}$ in Eq.~(\ref{Lag}) contribute to the dynamics. On the anisotropic background we are 
studying here, the $\tilde{{\cal L}}_5^{\rm N}$ term 
does not contribute to the dynamics at all.

For the matter sector, we consider 
a perfect fluid in terms of the k-essence description, i.e., 
${\cal L}_M=P(Z)$, where $Z=-g^{\mu \nu}
\partial_{\mu}\chi \partial_{\nu}\chi/2$ is 
the kinetic term of a scalar field $\chi$ \cite{kes}.
Then, the matter Lagrangian $L_M=\sqrt{-g}\,{\cal L}_M$, 
which corresponds to the last term of Eq.~(\ref{Lag}), reads
\be
L_M=Ne^{3\alpha}P(Z(N))\,, 
\ee
where $Z(N)=\dot{\chi}^2/(2N^2)$.
Varying $L_M$ with respect to $\chi$ and setting $N=1$ 
at the end, we obtain the continuity equation
\be
\dot{\rho}_M+3\dot{\alpha}\,(\rho_M+P_M)=0\,,
\ee
where $\rho_M$ and $P_M$ are given by 
\be
\rho_M=2ZP_{,Z}-P\,,\qquad 
P_M=P\,,
\ee
which correspond to the energy density and 
the pressure, respectively.
Note that $\rho_M$ and $P_M$ appear in
the background equations of motion derived 
by the variations of $N$ and $\alpha$, respectively. 

The Lagrangian (\ref{Lag}) contains the time-derivatives 
of $\alpha, \sigma, \phi,v, \chi$ up to first order, so the 
resulting equations of motion for these variables 
remain of second order.
If we compute the determinant ${\cal D}$ of the 
$6\times6$ Hessian matrix 
\be
H_L^{\mu \nu}=\frac{\partial^2 L}
{\partial \dot{\cal O}_{\mu} 
\partial \dot{\cal O}_{\nu}}\,,
\ee
where ${\cal O}=(N(t),\alpha(t), \sigma(t), \phi(t), v(t), 
\chi(t))$, it follows that ${\cal D}=0$ due to the absence of 
time derivatives of $N$ in Eq.~(\ref{Lag}). 
This suggests the existence of a constraint  that forbids 
the propagation of an additional ghostly DOF on that background.
In fact, variation of (\ref{Lag}) with respect to $N$ 
leads to the constraint equation
\be
\frac{\partial L}{\partial N}=0\,.
\label{Hamicon1}
\ee
Defining the conjugate momentum $\Pi^{\mu}=
\partial L/\partial \dot{{\cal O}}_{\mu}$, 
the Hamiltonian of the system is given by 
${\cal H}=\Pi^{\mu} \dot{\cal O}_{\mu}-L$. 
Introducing the isotropic expansion rate $H$ 
and the anisotropic  expansion rate $\Sigma$, as
\be
H \equiv \frac{\dot{\alpha}}{N}\,,\qquad 
\Sigma \equiv \frac{\dot{\sigma}}{N}\,,
\ee
the Hamiltonian reads 
\be
{\cal H}=e^{3\alpha} \left[N \rho_M 
-C_1(H+\Sigma) \dot{\phi}-C_2 
\frac{\dot{v}^2}{N}-C_3 \dot{v}-NC_4 \right]\,,
\label{Hami}
\ee
where 
\ba
&&
C_{1}=4 \phi v^{2} 
\left[ f_{4}+3 \phi ( H+\Sigma ) f_{5} \right]\,,\label{C1} \\
&&
C_{2}=-\frac12+g_{4}+4 \phi ( H+\Sigma ) g_{5}-3 
 ( H+\Sigma ) ^{2} \left[ G_{6}+\phi^{2} ( G_{6,X
}+2 f_{6} ) \right]\,,\\
&&
C_{3}=2 v \left[ ( H -2 \Sigma ) C_{2}-2  ( H+\Sigma ) ( G_{4,X}+\phi^{2}f_{4} ) 
 -\phi ( H+\Sigma ) ^{2} ( 6\phi^{2}f_{5}-G_{5,X} ) \right]
\,,\\
&&
C_{4}=G_{2}+6 ( H^2-\Sigma^2 ) 
 \left[ G_{4}-\phi^{2}G_{4,X}-\phi^{2} ( \phi^{2}-v^{2} ) f_{4} \right] 
 +2 \phi^{3} ( H -2 \Sigma ) ( H+\Sigma )^{2} \left[ G_{5,X}-6 ( \phi^{2}-v^{2} ) f_{5} \right] \nonumber \\
&&\hspace{1cm}
+v^{2} ( H -2 \Sigma )^{2}C_{2}\,.
\label{C4}
\ea
The Hamiltonian (\ref{Hami}) is related to the quantity
$\partial L/\partial N$, as
\be
\frac{\partial L}{\partial N}=-\frac{{\cal H}}{N}\,,
\label{LHN}
\ee
so the constraint (\ref{Hamicon1}) translates to 
\be
{\cal H}=0\,.
\ee
This means that the above system is not plagued by 
the Ostrogradski instability associated with the Hamiltonian 
unbounded from below \cite{Ostro}. 
Since the coefficient $C_1$ depends on the two 
functions $f_4$ and $f_5$, the derivative term 
$C_1(H+\Sigma)\dot{\phi}$ in Eq.~(\ref{Hami}) arises 
outside the domain of second-order generalized Proca 
theories. However, the Lagrangian densities ${\cal L}_4^{\rm N}$ 
and ${\cal L}_5^{\rm N}$ do not give rise to the appearance 
of an additional dangerous DOF related to the Ostrogradski ghost. 
Thus, the beyond-generalized theories remain healthy not only on 
the isotropic FLRW background \cite{HKT} but also on the 
anisotropic cosmological background.

Varying the Lagrangian (\ref{Lag}) with respect to 
$N$, $\alpha$, $\sigma$, $\phi$, $v$, respectively, and 
setting $N=1$ at the end, we obtain the dynamical 
equations of motion
\ba
&&
C_1(H+\Sigma)\dot{\phi}+C_2\dot{v}^2+C_3\dot{v}+C_4=\rho_M\,,\label{eqN}\\
&&
C_1 \ddot{\phi}+\left(C_5 \dot{v}+C_6 \right) \ddot{v} 
+\left(C_7 \dot{\phi}+C_8 \dot{v}^2+C_{9}\dot{v}+C_{10}\right) \dot{H}
+\left(C_7\dot{\phi}+C_8\dot{v}^2+C_{11}\dot{v}+C_{12}\right) \dot{\Sigma}\notag\\
&&
+C_{13}\dot{\phi}^2+\left(C_{14}\dot{v}^2+C_{15}\dot{v}+C_{16}\right) \dot{\phi}
+C_{17}\dot{v}^3+C_{18}\dot{v}^2+C_{19}\dot{v}+C_{20}=-3P_M\,,\label{eqa}\\
&&
\frac{d}{dt}\left[e^{3 \alpha} \left( C_1 \dot{\phi}
+\frac{C_5}{2}\dot{v}^2+C_{21}\dot{v}+C_{22}\right)\right]=0\,,\label{eqs}\\
&&
C_1 (\dot{H}+\dot{\Sigma})+D_{1}\dot{v}^2+D_{2}\dot{v}+D_{3}=0\,,\label{eqp}\\
&&
D_{4}\ddot{v}-(C_5 \dot{v}+C_6) \dot{H}-(C_5\dot{v}+C_{21}) \dot{\Sigma}
+(2D_{1}\dot{v}+D_{2}) \dot{\phi}+D_{5}\dot{v}^2+3 H D_{4} \dot{v}+D_{6}=0\,,\label{eqv} 
\ea
where the coefficients $C_{5,\cdots,22}$ and 
$D_{1,\cdots,6}$ are 
given in Appendix A.
The terms containing  $C_1$ appear in the beyond-generalized 
Proca theories. In this case, the dynamical evolution 
of the spatial vector component $v$ is different 
from that in the generalized Proca theories.
In Sec.~\ref{numesec} we shall discuss the cosmological dynamics 
in both generalized and beyond-generalized Proca theories in detail.

\section{Analytic solutions to the spatial anisotropy}
\label{anasec}

Let us analytically estimate the evolution of the anisotropic 
expansion rate $\Sigma$ and the spatial vector component $v$. 
To recover the expansion history close to 
that of GR in the early cosmological epoch, 
we consider the function 
\be
G_4(X)=\frac{M_{\rm pl}^2}{2}+\tilde{G}_4(X)\,,
\ee
where $M_{\rm pl}$ is the reduced Planck mass, 
and $\tilde{G}_4(X)$ is a function of $X$.
We write Eqs.~(\ref{eqN}) and (\ref{eqa}) 
in the following forms
\ba
& & 3M_{\rm pl}^2 H^2=\rho_M+\rho_{\rm DE}\,,
\label{Fri1} \\
& & M_{\rm pl}^2 \left( 2\dot{H}+3H^2 \right)=
-P_M-P_{\rm DE}\,,
\label{Fri2}
\ea
where $\rho_{\rm DE}$ and $P_{\rm DE}$ correspond 
to the energy density and the pressure of the ``dark'' 
component originating from the vector field, respectively.
Introducing the density parameters 
\be
\Omega_{M}=\frac{\rho_M}{3M_{\rm pl}^2 H^2}\,,
\qquad
\Omega_{\rm DE}=\frac{\rho_{\rm DE}}
{3M_{\rm pl}^2 H^2}\,,
\ee
we obtain the constraint 
$\Omega_M+\Omega_{\rm DE}=1$ from Eq.~(\ref{Fri1}).
We also define the effective equation of state $w_{\rm eff}$ 
and the dark energy equation of state $w_{\rm DE}$, as
\ba
w_{\rm eff}=-1-\frac{2\dot{H}}{3H^2}\,,\qquad
w_{\rm DE}=\frac{P_{\rm DE}}{\rho_{\rm DE}}\,.
\label{eqstate}
\ea
For the matter sector, we take into account radiation (labelled 
by ``$r$'') and non-relativistic matter (labelled by ``$m$''), 
such that $\rho_M=\rho_r+\rho_m$ and $P_M=\rho_r/3$.
Along the line of Refs.~\cite{cosmo,Geff}, 
the existence of a late-time de Sitter solution with 
$w_{\rm DE}=-1$ is assumed for the analytic estimation 
in Sec.~\ref{deSittersec}. In Sec.~\ref{numesec} we will present a 
concrete dark energy model with a de Sitter attractor.
The background expansion history is given by the cosmological sequence of 
radiation-dominated ($w_{\rm eff}=1/3$) $\to$ 
matter-dominated ($w_{\rm eff}=0$) $\to$ 
de Sitter ($w_{\rm eff}=-1$) epochs.

\subsection{Radiation and matter eras}
\label{anasecA}

During the radiation and deep matter eras, 
the dark energy density parameter $\Omega_{\rm DE}$ and 
the quantity $\epsilon_{P_{\rm DE}}=
P_{\rm DE}/(3M_{\rm pl}^2H^2)$ 
are much smaller than the order of unity. 
Except for the term $3M_{\rm pl}^2 H^2$, 
each term appearing on the l.h.s. of Eq.~(\ref{eqN}) 
can be assumed to be much smaller than 
$3M_{\rm pl}^2 H^2$, say, $|C_3 \dot{v}|/(3M_{\rm pl}^2H^2) \ll 1$. 
Since the first three terms inside the parenthesis 
of Eq.~(\ref{eqs}) are similar to 
those appearing in Eq.~(\ref{eqN}) (apart from 
the difference divided by $H$), we express them 
in the form 
\be
C_1 \dot{\phi}+\frac{C_5}{2}\dot{v}^2
+C_{21}\dot{v}=\epsilon M_{\rm pl}^2 H\,,
\ee
where $\epsilon$ is a dimensionless parameter.
Under the condition $|\Sigma| \ll H$, the 
dominant contribution to the term $C_{22}$ 
is given by 
\be
C_{22} \simeq -6M_{\rm pl}^2 \Sigma
-v^2 \left( 4H C_2 -\frac52 H^2 C_5 \right)\,.
\ee
Then, the equation of motion (\ref{eqs}) 
for $\Sigma$ reads
\be
\frac{d}{dt} \left[ a^3 \left\{ \Sigma-\frac16 \epsilon H
+\frac{v^2}{12M_{\rm pl}^2} \left( 8H C_2
-5H^2 C_5 \right)  \right\} \right]=0\,.
\label{Sigeq}
\ee

As long as the dark energy density is suppressed relative to 
the background fluid density, the parameter $\epsilon$ 
is much smaller than 1 in the early cosmological epoch. 
In Eq.~(\ref{eqN}) the term $-v^2H^2/2$ exists in the 
coefficient $C_4$, so we require the condition $v^2 \ll M_{\rm pl}^2$ 
in the early cosmological epoch.
Provided that the conditions 
\be
|\Sigma| \gg \left|\frac16 \epsilon H \right|\,,\qquad
|\Sigma| \gg \frac{v^2}{12M_{\rm pl}^2} \left| 8H C_2
-5H^2 C_5 \right|
\label{Sigcon}
\ee
are satisfied, $\Sigma$ evolves as 
\be
\Sigma \propto a^{-3}\,.
\label{Sigma}
\ee
We define the ratio between $\Sigma$ and $H$, as
\be
r_\Sigma \equiv \frac{\Sigma}{H}\,.
\ee
When $\Sigma$ decreases as Eq.~(\ref{Sigma}), the evolution of 
$r_{\Sigma}$ during the radiation and matter eras is given, 
respectively, by 
\ba
&&
r_{\Sigma} \propto a^{-1}\qquad({\rm radiation~era})\,,
\label{Sigrad}\\
&&
r_{\Sigma} \propto a^{-3/2}\qquad({\rm matter~era}).
\ea

If $v$ is not very small and the second condition 
of Eq.~(\ref{Sigcon}) does not hold, it happens that the anisotropy 
is sustained by $v$ in such a way that 
$\Sigma$ balances the term containing $v^2$ 
in Eq.~(\ref{Sigeq}). In Sec.~\ref{numesec} we shall study 
the evolution of $\Sigma$ in concrete dark energy models 
and show that, depending on model parameters and 
initial conditions, the ratio $r_{\Sigma}$ can stay 
constant in the radiation-dominated epoch.
This means that there are cases in which $r_{\Sigma}$ 
does not necessarily decrease as Eq.~(\ref{Sigrad}).

We also estimate the evolution of $v$ under the condition 
that the ratio $|r_{\Sigma}|$ is much smaller 
than 1. Neglecting the contributions of the terms $\Sigma$ 
and $\dot{\Sigma}$ to Eq.~(\ref{eqv}), the equation of 
motion for $v$ reads 
\be
\ddot{v}+\left( 3H +\frac{\alpha_1}{q_V} \right) \dot{v}
+\left( 2H^2+\dot{H}+\frac{\alpha_2}{q_V} \right)v 
\simeq 0\,, \label{vapeq}
\ee
where 
\ba
\hspace{-0.2cm}
q_V &=&
1-2g_4-4H\phi g_5
+2H^2 \left( G_6+\phi^2 G_{6,X}+2\phi^2 f_6 \right)\,,
\label{qV} \\
\alpha_1 &=&
4\dot{H} \left[ H G_6+\phi \{ (G_{6,X}+2f_6)H\phi
-g_5 \} \right] \nonumber \\
\hspace{-0.2cm}
& &
+2\dot{\phi} \left( \phi H 
[ H \{ 3G_{6,X}+4f_6+(G_{6,XX}+2f_{6,X})\phi^2 \}
-2\phi g_{5,X}]-g_{4,X}\phi-2Hg_{5} \right)\,,\\
\alpha_2 &=& G_{2,X}+6H^2 G_{4,X} +4\dot{H} 
(G_{4,X}+H^2 G_6)+\dot{v}^2 \left[ g_{4,X} 
-H \{ H G_{6,X}+\phi [(G_{6,XX}+2f_{6,X})H\phi
-2g_{5,X}] \} \right]+3H \phi G_{3,X} \nonumber \\
\hspace{-0.2cm}
& &
+2\dot{H} \phi \left[ 
2\phi \{ f_4+H(2H f_6 +HG_{6,X}+3\phi f_5)\}
-HG_{5,X}-2Hg_5  \right]+\phi H^2 [ 
\phi \{ 6G_{4,XX}+24f_4+\phi [30 Hf_5
-HG_{5,XX} \nonumber \\
\hspace{-0.2cm}
& & +6\phi (f_{4,X}+H\phi f_{5,X})] \}
-3HG_{5,X}]+H^2 v^2 ( H^2 G_{6,X}-g_{4,X}
+\phi [ H \{ (G_{6,XX}+2f_{6,X})H \phi
-6f_{5,X} \phi^2-2g_{5,X} \}
\nonumber \\
\hspace{-0.2cm}
& &
-6\phi f_{4,X} ])
+\dot{\phi} G_{3,X}+H\dot{\phi} \{ \phi 
[16f_4 -2g_{4,X}+4G_{4,XX}+8H^2f_6+6H^2G_{6,X} 
+\phi\{ 4\phi f_{4,X}
+H [30f_5-4g_{5,X}
\nonumber \\
\hspace{-0.2cm}
& &
-G_{5,XX}
+2(G_{6,XX}+2f_{6,X})H\phi+6\phi^2 f_{5,X}] \}
-2(2f_{4,X}+3H\phi f_{5,X})v^2] 
-HG_{5,X}-4H g_5 \}\,.
\ea
Note that we expressed the term $g_4$ in 
$\alpha_2$ by using $q_V$.
It is worthy of mentioning that the quantity $q_V$ is 
identical to the coefficient appearing in front of the 
kinetic term of vector perturbations on the isotropic 
background \cite{Geff,HKT}. To avoid the appearance of 
ghosts associated with vector perturbations, 
we require that $q_V>0$. Under the conditions 
\be
\left|\frac{\alpha_1}{q_V} \right| \ll H\,,\qquad
\left|\frac{\alpha_2}{q_V} \right| \ll H^2\,,
\label{alcon}
\ee
Eq.~(\ref{vapeq}) approximately reduces to 
\be
\ddot{v}+3H \dot{v}+\left( 2H^2+\dot{H} \right)v 
\simeq 0\,.
\label{vap2}
\ee
If the effective equation of state $w_{\rm eff}$ defined by 
Eq.~(\ref{eqstate}) is constant, the solution to Eq.~(\ref{vap2}) reads
\be
v=c_1 a^{-(1-3w_{\rm eff})/2}
+c_2a^{-1}\,,
\label{vsoma}
\ee
where $c_1$ and $c_2$ are integration constants. 
The evolution of $v$ in the early cosmological 
epoch is given by 
\ba
v &\propto& a^{0} \qquad \quad~({\rm for}~w_{\rm eff}=1/3),\label{vra}\\
v &\propto& a^{-1/2} \qquad ({\rm for}~w_{\rm eff}=0).
\label{vma}
\ea
Hence $v$ stays nearly constant during the radiation era, 
but it decreases in proportion to 
$t^{-1/3}$ during the matter era.

If the conditions (\ref{alcon}) are violated, 
the analytic solution (\ref{vsoma}) loses its validity.
In such cases, however, the large contribution from the 
spatial vector component to the background energy density 
can affect the successful cosmic expansion history. 
In Sec.~\ref{numesec} we shall study the dynamics of 
$v$ in concrete dark energy models.

\subsection{De Sitter fixed point}
\label{deSittersec}

In the absence of $v$ and $\Sigma$, it was shown in 
Refs.~\cite{cosmo,Geff} that there exist isotropic 
de Sitter solutions with constant $\phi$ and $H$ for the functions $G_{2,3,4,5}$ containing the power-law term in $X$.
Here, we would like to find other de Sitter fixed points at which 
$v,\Sigma$ as well as $\phi,H$ are non-zero constants. 
To discuss the stability of solutions, we shall keep the time derivative $\dot{\Sigma}$ in Eqs.~(\ref{eqs}) and the derivative terms $\ddot{v}, \dot{v}$ in Eq.~(\ref{eqv}), 
while dealing with $H$ and $\phi$ as constants.
Setting $\dot{v}=0, \ddot{v}=0$ in Eq.~(\ref{eqs}), it follows that 
\be
q_1 \dot{\Sigma}+q_2 \Sigma 
\simeq 3q_3H^2 v^2\,,
\label{Sigeq2}
\ee
where 
\ba
\hspace{-0.7cm}
q_1 &=& 6G_4 -6\phi^2 G_{4,X}+3\phi^3 
\left[ (G_{5,X}-6\phi^2 f_5 )(H+2\Sigma)-2\phi f_4 \right]
+v^2 [ 2-4g_4-3G_6(H^2-4H\Sigma-8\Sigma^2) \nonumber \\
\hspace{-0.7cm}
& &+3\phi \{ \phi [2f_4-H(HG_{6,X}-6\phi f_5
+2Hf_6)+4\Sigma (HG_{6,X}+3\phi f_5+2Hf_6)
+8\Sigma^2(G_{6,X}+2f_6)]-8g_5 \Sigma \}] 
\,, \\
\hspace{-0.7cm}
q_2 &=& 3Hq_1-3H\Sigma [3\phi^3 G_{5,X}
-18\phi^3 (\phi^2-v^2)f_5
+2v^2 \{[G_6+(G_{6,X}+2f_6)\phi^2](3H+8\Sigma)-6\phi g_5\}]\,,\label{q2} \\
\hspace{-0.7cm}
q_3 &=& 1-2g_4-3H\phi g_5
+H^2 \left( G_6+\phi^2 G_{6,X}+2\phi^2f_6 \right)\,.
\ea
Setting $\dot{\Sigma}=0$ in Eq.~(\ref{eqv}), keeping the 
terms linear in $\dot{v}, \ddot{v}$, and   
using Eq.~(\ref{eqp}), the equation  of motion 
for $v$ reads
\be
D_4 \left( \ddot{v}+3H \dot{v} \right)
+\frac{2(H+\Sigma)}{\phi} \left( H\phi q_V
+q_4-q_5v^2 \right) v \simeq 0\,,
\label{veqap}
\ee
where $q_V$ is of the same form as Eq.~(\ref{qV}), and 
\ba
\hspace{-0.7cm}
q_4 &=& \Sigma [ 18\phi^4 (H+\Sigma)f_5-2\phi \{
1-2g_4-3G_{4,X}+\Sigma (3H+2\Sigma) G_6 \}
-\phi^3 \{2\Sigma (3H+2\Sigma)(G_{6,X}+2f_6)-6f_4 \} 
\nonumber \\
\hspace{-0.7cm}
& &-\phi^2 \{3(H+\Sigma)G_{5,X}
-4(H+2\Sigma)g_5 \}]\,, \nonumber \\
\hspace{-0.7cm} 
q_5 &=&
H^2 \left( H\phi G_{6,X}+2H \phi f_6-g_5 \right)
+\Sigma [ 18(H+\Sigma)\phi^2 f_5+4(H-\Sigma)g_5
+6\phi f_4 -\phi(G_{6,X}+2f_6)(3H^2-4\Sigma^2) ].
\ea

On using Eqs.~(\ref{Sigeq2}) and (\ref{veqap}), 
we search for fixed points with constant values of $v$ and $\Sigma$ for $q_2 \neq 0$. 
One of them is the isotropic point characterized by 
\be
v=0\,,\qquad \Sigma=0\,.
\label{isofixed}
\ee
The other fixed points, which exist for $q_5 \neq 0$,  
obey the relations 
\ba
v^2 &=& \frac{H \phi q_V+q_4}{q_5}\,,\label{vfixed}\\
\Sigma &=&\frac{3q_3H^2 (H \phi q_V+q_4)}
{q_2 q_5}\,. \label{Sigfixed}
\ea
Not only the quantities $q_2, q_4, q_5$ contain $\Sigma$, 
but also the quantities $q_2,q_3,q_4,q_5,q_V$ depend on $v$ 
through the dependence of functions $G_{4,5,6}, f_{4,5,6}, g_{4,5}$ 
with respect to $X=\phi^2/2-v^2/2$. 
This means that, unless the functional forms 
of $G_4$ etc are specified, we cannot explicitly 
solve Eqs.~(\ref{vfixed}) and (\ref{Sigfixed}) for 
$v$ and $\Sigma$.
 
We are looking for the fixed points relevant to the late-time 
cosmic acceleration with a non-vanishing anisotropic expansion 
rate $\Sigma$. We assume that $\Sigma$ is much smaller than 
$H$ on the de Sitter solution, i.e.,  
\be
|r_{\Sigma}| \ll 1\,.
\label{dscon1}
\ee
Moreover, we focus on the case in which the condition 
\be
v^2 \ll \phi^2
\label{dscon2}
\ee
is satisfied on the de Sitter solution.
As we will study the dynamics of the vector field 
for concrete dark energy models in Sec.~\ref{numesec},  
the temporal component $\phi$ tends to grow toward 
the de Sitter fixed point in cosmologically viable cases. 
Meanwhile, as estimated by Eq.~(\ref{vma}),  
the spatial component $v$ typically decreases 
during the matter-dominated epoch.
Hence, even if $v$ is of the same order as $\phi$ in 
the radiation era, the condition (\ref{dscon2}) 
usually holds around the onset of cosmic acceleration. 
In what follows, we employ the approximation that the functions $G_{4,5,6}, f_{4,5,6}, g_{4,5}$ do not depend on $v$ by dealing with the 
kinetic term as $X \simeq \phi^2/2$.

Under the approximation (\ref{dscon1}) we ignore the terms 
containing $\Sigma$ in Eq.~(\ref{q2}), in which case 
$q_2 \simeq 3Hq_1$. 
Moreover we also neglect the $\Sigma$-dependent terms 
in Eq.~(\ref{veqap}), in which case $D_4 \simeq q_V$.
Then Eqs.~(\ref{Sigeq2}) and (\ref{veqap}) reduce, 
respectively,  to 
\ba
& & q_1 \left( \dot{\Sigma}+3H \Sigma \right) 
\simeq 3q_3 H^2 v^2\,,
\label{vsigap2} \\
& & q_V \left( \ddot{v}+3H \dot{v}+2H^2v \right) 
\simeq \frac{2H^3}{\phi} {\cal A}_V v^3\,,
\label{veqap2}
\ea
where 
\be
{\cal A}_V \equiv 
H\phi \,G_{6,X}+2H \phi f_6-g_5\,.
\label{AV}
\ee

Let us first consider the theories characterized by  
\be
g_5=0\,,\qquad G_6=0\,,\qquad
f_6=0\,,
\label{theory1}
\ee
in which case ${\cal A}_V=0$ with $q_V=1-2g_4$. 
Provided that $g_4 \neq 1/2$, Eq.~(\ref{veqap2}) reduces to 
\be
\ddot{v}+3H \dot{v}+2H^2v=0\,,
\ee
whose solution is given by 
\be
v=c_3 a^{-1}+c_4 a^{-2}\,,
\ee
where $c_3$ and $c_4$ are integration constants.
Hence the spatial component $v$ exponentially 
decreases toward 0.
Since the r.h.s. of Eq.~(\ref{vsigap2}) approaches 0, 
the equation for $\Sigma$ reduces to 
$\dot{\Sigma}+3H\Sigma \simeq 0$.
This means that the anisotropic expansion rate 
decays as $\Sigma \propto a^{-3}$.
If $g_4=1/2$ then we have $q_3=1-2g_4=0$, 
so the r.h.s. of Eq.~(\ref{vsigap2}) vanishes. 
These discussions show that, for the theories 
given by the functions (\ref{theory1}), both $v$ 
and $\Sigma$ decrease 
toward the isotropic fixed point (\ref{isofixed}).

We proceed to the theories in which the terms 
$g_5,G_6,f_6$ are present, such that 
\be
{\cal A}_V \neq 0\,.
\ee
In this case, besides the isotropic point (\ref{isofixed}), 
there exist other fixed points satisfying Eqs.~(\ref{vfixed}) 
and (\ref{Sigfixed}). Ignoring the $v$ dependence in 
the functions $G_{4,5,6}, f_{4,5,6}, g_{4,5}$ under the 
condition (\ref{dscon2}), the latter fixed points 
correspond to $v=v_c$ and $\Sigma=\Sigma_c$, where
\be
v_c=\pm \sqrt{\frac{\phi\,q_V}{H{\cal A}_V}}\,,\qquad 
\Sigma_c=\frac{q_3q_V \phi}{\tilde{q}_1{\cal A}_V}\,,
\label{vc}
\ee
with
\be
\tilde{q}_1=6G_4 -6\phi^2 G_{4,X}+3\phi^3 
\left[ (G_{5,X}-6\phi^2 f_5 )H-2\phi f_4 \right]
+v_c^2 \left[ 2-4g_4-3H^2 G_6+3\phi^2 \{ 2f_4 
-H(H G_{6,X}-6\phi f_5+2H f_6) \} \right].
\ee
Existence of the fixed point (\ref{vc}) requires that 
$\phi\,q_V/(H{\cal A}_V)>0$.
Considering a linear perturbation $\delta v$ around 
$v=v_c$ in Eq.~(\ref{veqap2}), 
it follows that 
\be
\ddot{\delta v}+3H \dot{\delta v}-4H^2 \delta v=0\,,
\ee
whose solution is given by 
\be
\delta v=\tilde{c}_3 a+
\tilde{c}_4 a^{-4}\,,
\ee
where $\tilde{c}_3$ and $\tilde{c}_4$ are 
integration constants.
Since $\delta v$ exponentially grows in time, 
the fixed point (\ref{vc}) is not stable.
On the other hand, the perturbation around the fixed point 
$v=0$ obeys the differential equation 
$\ddot{\delta v}+3H \dot{\delta v}+2H^2 \delta v=0$, 
so $\delta v$ exponentially decreases toward 0.
Then, the solutions finally approach the fixed point 
$v=0$ rather than $v=v_c$. On using Eq.~(\ref{vsigap2}), 
$\Sigma$ also decreases as $\propto a^{-3}$ toward 
the isotropic point $\Sigma=0$.
The above discussion shows that, under the conditions 
(\ref{dscon1}) and (\ref{dscon2}), 
the anisotropic hair with a non-vanishing 
$\Sigma$ does not survive in general  
on the de Sitter background. 
This is consistent with the Wald's 
cosmic conjecture \cite{Wald}.

\section{Anisotropic cosmological dynamics 
in concrete dark energy models}
\label{numesec}

We study the anisotropic cosmological dynamics for 
a class of dark energy models in the framework of 
massive vector theories.
We consider the functions $G_{2,3,4,5}$ containing 
the power-law dependence of $X$, such that 
\be
G_2(X)=b_2 X^{p_2}\,,\qquad
G_3(X)=b_3X^{p_3}\,,\qquad
G_4(X)=\frac{M_{\rm pl}^2}{2}+b_4X^{p_4}\,,\qquad
G_5(X)=b_5X^{p_5}\,,
\label{G2345}
\ee
where $M_{\rm pl}$ is the reduced Planck mass,
$b_{2,3,4,5}$ and $p_{2,3,4,5}$ are constants.
In the isotropic context ($v=0=\Sigma$),
the simple solution $\phi^p \propto H^{-1}$ 
can be realized for the powers \cite{cosmo,Geff}
\be
p_3=\frac12 \left( p+2p_2-1 \right)\,,\qquad
p_4=p+p_2\,,\qquad
p_5=\frac12 \left( 3p+2p_2-1 \right)\,,
\label{p345}
\ee
which accommodate the vector Galileon \cite{Heisenberg} 
as a specific case ($p_2=1$ and $p=1$).
Provided that $p>0$, the temporal component $\phi$ 
grows with the decrease of $H$.
Finally, the solutions approach the de Sitter fixed 
point characterized by constant $\phi$ and $H$.
On the FLRW background the $G_6(X)$ term does not 
contribute to the dynamics \cite{Geff}.

In the isotropic setting, the dark energy density 
$\rho_{\rm DE}^{\rm (iso)}$ and the pressure 
$P_{\rm DE}^{\rm (iso)}$ originate from the 
temporal vector component $\phi$. 
In this case, the dark energy equation of state 
$w_{\rm DE}^{\rm (iso)}=P_{\rm DE}^{\rm (iso)}/
\rho_{\rm DE}^{\rm (iso)}$
is analytically known as \cite{cosmo}
\be
w_{\rm DE}^{\rm (iso)}=
-\frac{3(1+s)+s\,\Omega_r}
{3(1+s\,\Omega_{\rm DE})}\,,
\label{wdeiso}
\ee
where $s \equiv p_2/p$, and $\Omega_r=\rho_r
/(3M_{\rm pl}^2H^2)$ is the radiation density parameter.
The evolution of $w_{\rm DE}^{\rm (iso)}$ is given by 
$w_{\rm DE}^{\rm (iso)} \simeq -1-4s/3$ 
in the radiation-dominated epoch 
($\Omega_r \simeq 1, \Omega_{\rm DE} \simeq 0$), 
$w_{\rm DE}^{\rm (iso)} \simeq -1-s$ 
in the matter-dominated epoch 
($\Omega_r \simeq 0, \Omega_{\rm DE} \simeq 0$), 
and $w_{\rm DE}^{\rm (iso)} \simeq -1$ 
during the dark energy dominance ($\Omega_r \simeq 0, 
\Omega_{\rm DE} \simeq 1$). 
For $s>0$ the phantom dark energy equation of state  
can be realized during the radiation and matter eras, 
but the parameter $s$ is constrained to be $s \leq 0.36$ 
at 95 \%\,confidence level for the compatibility 
with observations \cite{DeFe}.

In the anisotropic setting, there are contributions 
to the background equations of motion (\ref{Fri1}) and (\ref{Fri2}) 
originating from $v$ and $\Sigma$ besides $\rho_{\rm DE}^{\rm (iso)}$ 
and $P_{\rm DE}^{\rm (iso)}$. 
As estimated by Eq.~(\ref{vra}), 
let us consider the case in which $v$ stays nearly constant 
during the radiation era with $|r_{\Sigma}| \ll 1$.
Then, the dominant contribution to the dark energy 
density $\rho_{\rm DE}$ arising from $v$ is the 
term $C_2 v^2 H^2$ in Eq.~(\ref{C4}).
Moreover, the contribution 
$3M_{\rm pl}^2 \Sigma^2$ in $C_4$ 
cannot be necessarily neglected relative to 
$\rho_{\rm DE}^{(\rm iso)}$ even for $|r_{\Sigma}| \ll 1$. 
Similarly, the terms $3C_2 v^2H^2$ and 
$9M_{\rm pl}^2 \Sigma^2$ in $C_{20}$ and 
the $v^2$-dependent term in $C_{10}$ give rise 
to the dominant contribution to the pressure $P_{\rm DE}$.
Then, during the radiation era, we can estimate 
$\rho_{\rm DE}$ and $P_{\rm DE}$, as
\ba
\rho_{\rm DE} &\simeq& \rho_{\rm DE}^{(\rm iso)}
-{\cal C}_2 v^2 H^2+3M_{\rm pl}^2 \Sigma^2\,,
\label{rhoDE} \\
P_{\rm DE} &\simeq& P_{\rm DE}^{(\rm iso)}
+v^2 H^2 \left( {\cal C}_2-\frac23 {\cal C}_{10} 
\right)+3M_{\rm pl}^2 \Sigma^2\,,
\label{PDE}
\ea
where 
\ba
{\cal C}_2 
&=&-\frac12 q_V+2H\phi g_5-2H^2 \left( G_6
+\phi^2 G_{6,X}+2\phi^2 f_6 \right)\,,
\label{calC2} \\
{\cal C}_{10}
&=&-q_V+8H\phi g_5-10H^2 \left( G_6
+\phi^2 G_{6,X}+2\phi^2 f_6 \right)
+12\phi^2 \left( f_4+3H \phi f_5 \right)\,.
\label{calC10}
\ea

As we go back to the past, the isotropic contributions 
to Eqs.~(\ref{rhoDE}) and (\ref{PDE}) get
smaller \cite{cosmo}, whereas the terms containing 
$v$ and $\Sigma$ tend to be larger.
Provided that the two conditions 
\be
|{\cal C}_{2}|\frac{v^2}{M_{\rm pl}^2} \gg 
r_{\Sigma}^2\,,\qquad 
|{\cal C}_{2}| v^2 H^2 \gg \rho_{\rm DE}^{(\rm iso)}
\label{vin}
\ee
are satisfied with $|P_{\rm DE}^{(\rm iso)}|$ and $|{\cal C}_{10}|$ same order as 
$\rho_{\rm DE}^{(\rm iso)}$ and $|{\cal C}_{2}|$, respectively, the dark energy 
equation of state $w_{\rm DE}=P_{\rm DE}/\rho_{\rm DE}$ is 
approximately given by 
\be
w_{\rm DE} \simeq -1+\frac{2{\cal C}_{10}}
{3{\cal C}_2}\,.
\label{wdees1}
\ee
If the terms containing $g_5, G_6, f_{4,5,6}$ 
provide the sub-dominant contributions to 
${\cal C}_2$ and ${\cal C}_{10}$, we have that 
${\cal C}_{10} \simeq 2{\cal C}_2$ 
and hence Eq.~(\ref{wdees1}) reduces 
to $w_{\rm DE} \simeq 1/3$.
In this case, the spatial vector component $v$ 
works as a dark radiation. 

On the other hand, if the anisotropic expansion rate 
is large such that the conditions
\be
r_{\Sigma}^2 \gg |{\cal C}_{2}|\frac{v^2}{M_{\rm pl}^2}\,,\qquad
r_{\Sigma}^2 \gg \frac{\rho_{\rm DE}^{(\rm iso)}}
{3M_{\rm pl}^2H^2}
\label{vin2}
\ee
are satisfied, it follows that
\be
w_{\rm DE} \simeq 1\,.
\ee
The above estimations are valid during the radiation era 
in which the isotropic contributions to Eqs.~(\ref{rhoDE}) 
and (\ref{PDE}) are smaller than the terms 
containing $v$ and $\Sigma$, but after the temporal 
vector component $\phi$ dominates the dynamics, the evolution 
of $w_{\rm DE}$ is described by $w_{\rm DE}^{(\rm iso)}$ 
in Eq.~(\ref{wdeiso}).

In the following we shall study the cosmological 
dynamics in the three models: 
(A) $G_{2,3,4,5} \neq 0, G_6=0, g_4 \neq 0, g_5=0, f_{4,5,6}=0$, 
(B) $G_{2,3,4,5} \neq 0, G_6 \neq 0, g_4=0, g_5 \neq 0, f_{4,5,6}=0$,
and 
(C) $f_{4,5,6} \neq 0$ with all the other functions
$G_{2,3,4,5,6}, g_{4,5}$ non-vanishing.
For $G_2(X)$, we take the functional form 
\be
G_2(X)=-m^2 X\,,
\label{G2}
\ee
where $m$ is a constant having a dimension of mass. 
Since $p_2=1$ in this case, the parameter $s$ in Eq.~(\ref{wdeiso}) 
is equivalent to $s=1/p$.
The minus sign of $G_2(X)$ is chosen to avoid the tensor ghost 
and instability at the de Sitter fixed point \cite{cosmo}. 
Since the effective mass squared of the vector field is 
$2H^2$ on the de Sitter solution \cite{cosmo,Geff}, 
the tachyonic instability is absent even for the choice (\ref{G2}). 
We shall consider the case in which the bare mass $m$ is 
of the order of the present Hubble parameter $H_0$ 
with the temporal component $\phi_0 \approx 
{\cal O}(M_{\rm pl})$, such that the quantity defined by 
\be
\xi \equiv \frac{H}{m} \left( \frac{\phi}{M_{\rm pl}} 
\right)^{p}
\label{xi}
\ee
is of the order of unity. 
If $\phi$ dominates over $v$ during the cosmological 
evolution, the quantity $\xi$ stays nearly constant. 
We also introduce the following dimensionless 
constants
\be
a_3=\frac{M_{\rm pl}^p b_3}
{2^{p_3}m}\,,\qquad
a_4=\frac{M_{\rm pl}^{2p} b_4}
{2^{p_4}}\,,\qquad
a_5=\frac{M_{\rm pl}^{3p}m b_5}
{2^{p_5}}\,,
\ee
for the numerical purpose.

The structure of Eqs.~(\ref{eqN})-(\ref{eqv}) is
different depending on whether the Lagrangian densities
(\ref{L4N}) and (\ref{L5N}) are present or not.
If $f_4=f_5=0$, then the terms containing $C_1$ in 
Eqs.~(\ref{eqN})-(\ref{eqp}) vanish identically.
In such cases, we take the time derivative of Eq.~(\ref{eqp}) 
and solve for $\ddot{v}$, $\dot{\phi}$, $\dot{H}$, and 
$\dot{\Sigma}$ by using Eqs.~(\ref{eqa}), (\ref{eqs}), 
and (\ref{eqv}).
Then the dynamical equations are expressed 
in the autonomous form
\be
Z {\bm x}={\bm y}\,,
\label{auto}
\ee
where ${\bm x}={}^t(\ddot{v},\dot{\phi},\dot{H},\dot{\Sigma})$, 
$Z$ and ${\bm y}$ are the $4 \times 4$ and $1 \times 4$ matrices, 
respectively, containing the dependence of $\dot{v},v, \phi,H, \Sigma$. 
Provided that the determinant of $Z$ does not vanish, 
we can solve Eq.~(\ref{auto}) for ${\bm x}$, as 
${\bm x}=Z^{-1}{\bm y}$.
As usual, the Friedmann equation (\ref{eqN}) 
can be used as a constraint equation.

In beyond-generalized Proca theories ($f_{4,5} \neq 0$), 
there are extra derivative terms related to the non-vanishing 
coefficient $C_1$. Then the structure of the dynamical system 
is different from the one discussed above.
In Sec.~\ref{bgsec} we shall separately study such cases.

\subsection{$G_{2,3,4,5} \neq 0, G_6=0, g_4 \neq 0, g_5=0, f_{4,5,6}=0$}
\label{secA}

In this case the quantity $q_V$ is equivalent to $1-2g_4$. 
Let us consider the model with constant $g_4$ with 
$q_V>0$, i.e., $g_4<1/2$.
In the radiation-dominated epoch, the dominant contribution 
to the Friedmann equation (\ref{eqN}) originating from 
$v$ corresponds to the energy 
density $\rho_{g_4}=-C_2H^2 v^2=q_VH^2 v^2/2$.
The density parameter associated with 
$\rho_{g_4}$ is given by 
\be
\Omega_{g_4}\equiv \frac{\rho_{g_4}}{3M_{\rm pl}^2 H^2}
=\frac{q_V}{6} \frac{v^2}{M_{\rm pl}^2}\,.
\ee
Provided that $v^2 \ll M_{\rm pl}^2$, $\Omega_{g_4}$ 
is much smaller than 1 for $q_V \lesssim 1$. 
In the following, we shall consider the case in which 
the condition $\Omega_{g_4} \ll 1$ is satisfied
in the deep radiation era with $v^2 \lesssim \phi^2$.

For the model under consideration we have 
$\alpha_1=0$ in Eq.~(\ref{vapeq}), whereas 
the quantity $\alpha_2$ reduces to 
\ba
\alpha_2
&=&-m^2+(\dot{\phi}+3H\phi)G_{3,X}+2(3H^2+2\dot{H})G_{4,X}
+2H\phi(2\dot{\phi}+3H\phi)G_{4,XX} \nonumber \\
& &-H (H \dot{\phi}+2\dot{H}\phi+3H^2 \phi)G_{5,X}
-H^2 \phi^2 (\dot{\phi}+H\phi) G_{5,XX}\,.
\label{alpha2}
\ea
As long as the condition $|\alpha_2/q_V| \ll H^2$ 
is satisfied in the early cosmological epoch, 
the solution to Eq.~(\ref{vapeq})
is given by Eq.~(\ref{vra}) in the radiation era and 
by Eq.~(\ref{vma}) in the matter era.

%%%%%%%%%%%%%%%%%%%%%%%%%%%%%%
\begin{figure}
\begin{center}
\includegraphics[height=3.4in,width=3.4in]{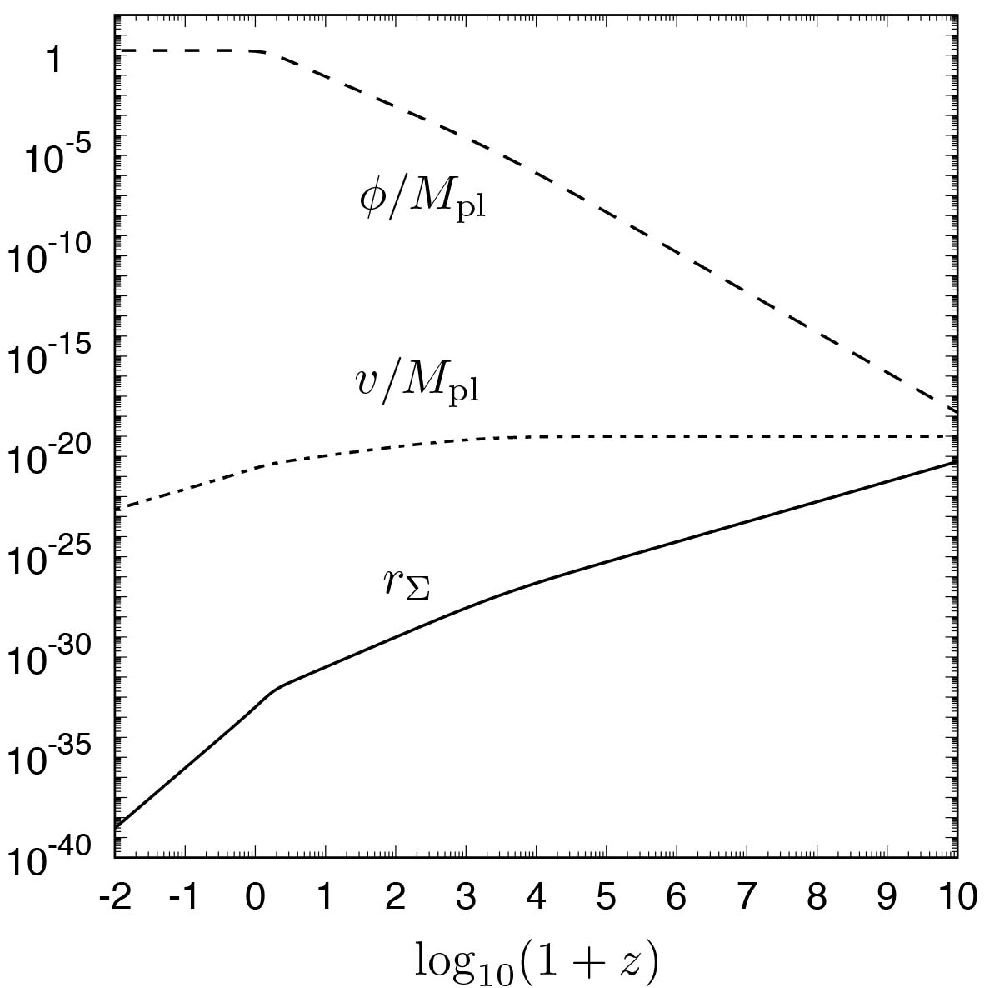}
\includegraphics[height=3.5in,width=3.4in]{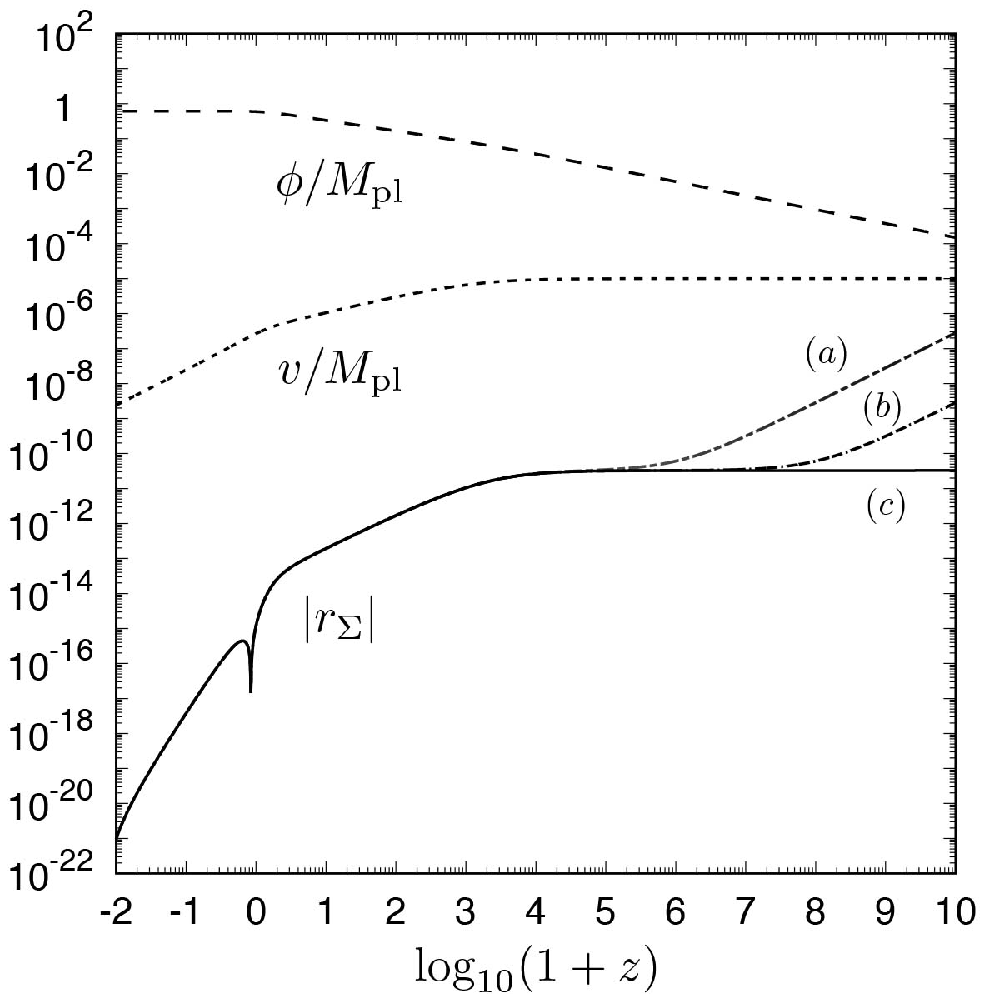}
\end{center}
\caption{\label{fig1}
Evolution of $\phi/M_{\rm pl}$, $v/M_{\rm pl}$, and $r_{\Sigma}$ 
in the model (A) with the parameters $a_4=0.01$, $a_5=0.05$, 
and $g_4=0.01$ for $p=1$ (left) and $p=5$ (right). 
We choose the initial conditions as 
$\phi/M_{\rm pl}=5.0\times10^{-19}$, 
$v/M_{\rm pl}=1.0\times10^{-19}$, 
$\dot{v}=0$,
$m\phi/(\sqrt{6} M_{\rm pl} H)=1.0\times10^{-37}$, 
$r_{\Sigma}=5.0\times10^{-20}$ and 
$1-\Omega_r=1.7\times10^{-7}$ 
at the redshift $z=1.8\times10^{10}$ (left), 
and $\phi/M_{\rm pl}=9.0\times10^{-5}$, 
$v/M_{\rm pl}=1.0\times10^{-5}$, 
$\dot{v}=0$,
$m\phi/(\sqrt{6} M_{\rm pl} H)=1.0\times10^{-23}$, 
and $1-\Omega_r=9.0\times10^{-8}$ 
at $z=3.5\times10^{10}$ (right). 
The cases (a), (b), (c) in the right panel correspond 
to the three different initial conditions: 
(a) $r_{\Sigma}=10^{-6}$, (b) $r_{\Sigma}=10^{-8}$, and 
(c) $r_{\Sigma}=3.3 \times 10^{-11}$ at $z=3.5\times10^{10}$.
The parameter $a_3$ is known from Eq.~(\ref{eqp}).
The present epoch ($z=0$) is identified by 
$\Omega_{\rm DE}=0.68$. }
\end{figure}
%%%%%%%%%%%%%%%%%%%%%%%%%%%%%%

If $v$ is nearly constant during the radiation domination, 
the terms containing $v^2$ in Eq.~(\ref{Sigeq}) affects
the evolution of $\Sigma$. Neglecting the contribution 
$\epsilon H/6$ relative to $\Sigma$ 
and using the fact that $C_2=-q_V/2$ and $C_5=0$, 
the solution to Eq.~(\ref{Sigeq}) is given by 
\be
r_{\Sigma} \simeq \frac{q_Vv^2}{3M_{\rm pl}^2}+
\frac{{\cal B}}{a^3 H}\,,
\label{rsigso}
\ee
where ${\cal B}$ is an integration constant.
The first term on the r.h.s. of Eq.~(\ref{rsigso})
stays constant, whereas the second term decreases as 
Eq.~(\ref{Sigrad}) in the radiation domination.
Unless the first contribution is extremely smaller than 
the second one around the beginning of the radiation era, 
the anisotropic expansion rate should approach
the value $r_{\Sigma} \simeq q_Vv^2/(3M_{\rm pl}^2)$.

In the left panel of Fig.~\ref{fig1} we show an example 
for the evolution of $v$, $\phi$, and $r_{\Sigma}$ in 
the case of vector Galileons ($p=1$) with $g_4=0.01$.
In this simulation we confirmed that the condition $|\alpha_2/q_V| \ll H^2$ 
is well satisfied before the onset of cosmic acceleration, 
so the analytic solution (\ref{vsoma}) to $v$ should be trustable.
In fact $v$ stays nearly constant during the radiation era, 
which is followed by 
the decrease of $v$ after the matter dominance.
In the left panel of Fig.~\ref{fig1} the temporal component $\phi$ 
grows as $\phi \propto H^{-1}$ toward the de Sitter attractor 
characterized by constant $\phi$, whose property is similar to 
the isotropic case studied in Ref.~\cite{cosmo}.

For $p=1$, the numerical simulation shows that $\Sigma$ 
decreases as $\propto a^{-3}$ during the radiation 
and matter eras. 
This reflects the fact that the initial 
value of $v$, which is at most of the same order as $\phi$, 
is very much smaller than $M_{\rm pl}$ ($v=10^{-19}M_{\rm pl}$ 
in the left panel of Fig.~\ref{fig1}), so the 
system enters the matter-dominated epoch before 
$r_{\Sigma}$ approaches the value $q_Vv^2/(3M_{\rm pl}^2)$ 
in the radiation era.
After the matter dominance, the solutions finally approach the 
isotropic fixed point characterized by $\Sigma=0$ and $v=0$.
Since the model under consideration belongs to the class 
(\ref{theory1}), there is no anisotropic de Sitter fixed point 
satisfying the conditions (\ref{dscon1}) 
and (\ref{dscon2}).

In the right panel of Fig.~\ref{fig1} we plot the evolution 
of $v,\phi, r_{\Sigma}$ for $p=5$ and $g_4=0.01$ 
with three different initial values of $r_{\Sigma}$.
The evolution of $\phi$ is similar to that in the isotropic case 
($\phi \propto H^{-1/p}$), so the variation of $\phi$ 
tends to be milder for larger $p~(>0)$. 
For $p=5$, it is then possible to choose larger initial values 
of $\phi$ and $v$ relative to those for $p=1$.
In the right panel of Fig.~\ref{fig1} the field $v$ stays nearly 
constant ($v \simeq 10^{-5}M_{\rm pl}$) during the radiation era, 
so the first term on the r.h.s. of Eq.~(\ref{rsigso}) 
can be estimated as $q_Vv^2/(3M_{\rm pl}^2) \simeq 3 \times 10^{-11}$.
We choose several different initial values of $r_{\Sigma}$ 
and find that the solutions temporally approach the value 
$r_{\Sigma}=q_Vv^2/(3M_{\rm pl}^2)$ in the radiation era, 
see the curves (a), (b), (c) in Fig.~\ref{fig1}.
This shows that, for the models allowing large initial 
values of $v$ (i.e., for greater $p$), 
the solution $r_{\Sigma}=q_Vv^2/(3M_{\rm pl}^2)$ 
corresponds to the temporal attractor in the 
radiation-dominated epoch.
After the onset of the matter domination, $r_{\Sigma}$ starts 
to decrease with the decrease of $v$.

As for the early evolution of $w_{\rm DE}$, we should 
notice that the quantities (\ref{calC2}) and (\ref{calC10}) 
reduce, respectively, to ${\cal C}_2=-q_V/2$ and 
${\cal C}_{10}=-q_V=2{\cal C}_2$. 
If the conditions (\ref{vin}) are satisfied during the radiation 
era, the dark energy equation of state (\ref{wdees1}) 
reduces to $w_{\rm DE} \simeq 1/3$. 
On the other hand, if the anisotropic expansion rate 
is initially large to satisfy the condition (\ref{vin2}), 
we have that $w_{\rm DE} \simeq 1$.

%%%%%%%%%%%%%%%%%%%%%%%%%%%%%%
\begin{figure}
\begin{center}
\includegraphics[height=3.4in,width=3.5in]{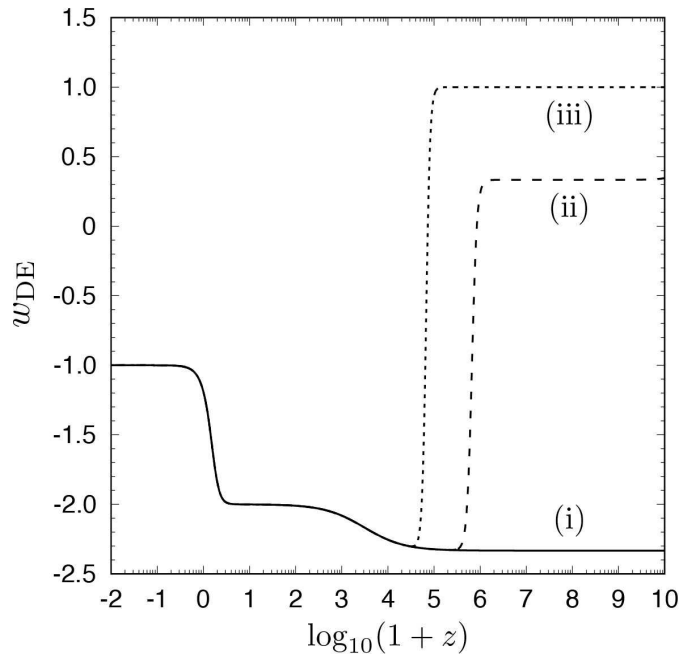}
\end{center}
\caption{\label{fig2}
Evolution of $w_{\rm DE}$ in the model (A) 
for $p=1$ with $a_4=0.01$, $a_5=0.05$, and $g_4=0.01$.
We choose the three different initial conditions: 
(i) $v=0$ and $r_{\Sigma}=0$, 
(ii) $v/M_{\rm pl}=1.0 \times 10^{-19}$ and 
$r_{\Sigma}=5.0\times10^{-20}$, and
(iii) $v/M_{\rm pl}=1.0\times10^{-19}$ and 
$r_{\Sigma}=1.0\times10^{-10}$ 
at $z=1.8\times10^{10}$. 
The initial conditions of other variables are 
the same as those used in the left panel of Fig.~\ref{fig1}.}
\end{figure}
%%%%%%%%%%%%%%%%%%%%%%%%%%%%%%

%%%%%%%%%%%%%%%%%%%%%%%%%%%%%%
\begin{figure}
\begin{center}
\includegraphics[height=3.4in,width=3.5in]{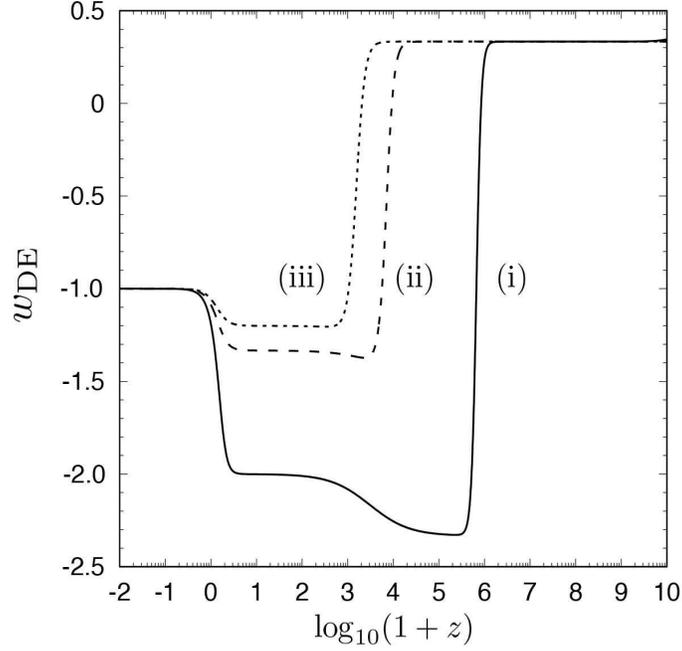}
\end{center}
\caption{\label{fig3}
Evolution of $w_{\rm DE}$ in the model (A) with 
$a_4=0.01$, $a_5=0.05$, and $g_4=0.01$ 
for three different cases: 
(i) $p=1$, (ii) $p=3$, and (iii) $p=5$.
The initial conditions for the cases (i) and (iii) are 
the same as those used in the left and right panels 
of Fig.~\ref{fig1}, while
for the case (ii) the initial conditions are chosen 
to be $\phi/M_{\rm pl}=9.0\times10^{-8}$, 
$v/M_{\rm pl}=5.0\times10^{-8}$, 
$\dot{v}=0$, $m\phi/(\sqrt{6} M_{\rm pl} H)=1.0\times10^{-28}$, 
$r_{\Sigma}=1.0\times10^{-12}$, and
$1-\Omega_r=1.1\times10^{-8}$ at $z=3.1\times10^{11}$. 
In these cases the conditions (\ref{vin}) are satisfied  
at $z=10^{10}$, so $w_{\rm DE}$ starts to evolve from 
the value $1/3$ in the radiation era.}
\end{figure}
%%%%%%%%%%%%%%%%%%%%%%%%%%%%%%

In Fig.~\ref{fig2} we plot the evolution of $w_{\rm DE}$ 
for $p=1$ with three different initial values of 
$v$ and $\Sigma$ at the redshift $z=10^{10}$.
In the case (i) we have chosen the isotropic value 
$v=0=\Sigma$, so $w_{\rm DE}$ evolves according to Eq.~(\ref{wdeiso}), i.e., $-7/3$ (radiation era) $\to$
$-2$ (matter era) $\to$ $-1$ (de Sitter epoch). 
In the case (ii) the initial conditions are the same as those used 
in the left panel of Fig.~\ref{fig1}. In this case the conditions (\ref{vin}) are satisfied in the deep radiation era, so that 
$w_{\rm DE} \simeq 1/3$.
After the contribution of $\phi$ to $w_{\rm DE}$ 
dominates over that of $v$, the evolution of $w_{\rm DE}$ 
is described by $w_{\rm DE}^{({\rm iso})}$.
In the numerical simulation of the case (ii), the approach of 
$w_{\rm DE}$ to $w_{\rm DE}^{({\rm iso})}$ occurs 
around the redshift $z=10^6$.
In the case (iii) of Fig.~\ref{fig2}, $\Sigma$ is initially large 
to fulfill the conditions (\ref{vin2}), so $w_{\rm DE}$ 
is close to 1.
In this case $\Sigma$ decreases in proportion to $a^{-3}$,
so the solutions finally enter the regime in which 
$w_{\rm DE} \simeq w_{\rm DE}^{(\rm iso)}$  
for $z \lesssim 10^5$.
Realization of the case (iii) requires that $\Sigma$ is 
of the order of $r_{\Sigma} \gtrsim 10^{-15}$ at $z=10^{10}$.

For $p=1$ the deviation of $w_{\rm DE}$ from $-1$ is 
significant during the matter era, so it is difficult for 
vector Galileons to be compatible with observations \cite{DeFe}.
However, this situation is different for larger $p$ 
(i.e., for smaller $s=1/p$).
In Fig.~\ref{fig3} we plot the evolution of $w_{\rm DE}$ 
for $p=1,3,5$ with the initial values of $v$ and $\Sigma$ 
obeying the conditions (\ref{vin}) at $z=10^{10}$. 
Hence $w_{\rm DE}$ starts to evolve from 
the value close to $1/3$, which is followed by 
the approach to the isotropic value 
$w_{\rm DE}^{({\rm iso})}=-1-1/p$ in the matter era.
For larger $p$ the evolution of $\phi$ during the radiation era 
is milder (as seen in Fig.~\ref{fig1}), 
so the difference between $v$ and $\phi$   
becomes less significant with the passage of time for the initial
values of $v$ same order as $\phi$. 
Then, for larger $p$, the approach of 
$w_{\rm DE}$ to $w_{\rm DE}^{({\rm iso})}$ tends to occur 
at the later cosmological epoch, but as long as $p={\cal O}(1)$, 
the transition redshift is much larger than 1.

In the regime where the ratio $r_{\Sigma}$ is close to
$q_V v^2/(3M_{\rm pl}^2)$, the last terms on the r.h.s. of 
Eqs.~(\ref{rhoDE}) and (\ref{PDE}) are about 
$v^2/M_{\rm pl}^2~(\ll 1)$ times as small as 
the second terms, so the first condition of Eq.~(\ref{vin}) 
is satisfied. During the radiation era in which the second 
condition of Eq.~(\ref{vin}) is fulfilled as well, 
the spatial vector component behaves as a dark radiation 
with $w_{\rm DE} \simeq 1/3$. 
In the case (iii) of Fig.~\ref{fig3}, we can confirm 
that $w_{\rm DE} \simeq 1/3$ by the time 
at which $r_{\Sigma}$ starts to decrease from the value 
$q_V v^2/(3M_{\rm pl}^2)$ around $z \approx 10^4$ 
(see the right panel of Fig.~\ref{fig1}).
Note that, in the gauge-quintessence scenario studied 
in Ref.~\cite{gaugequ}, non-Abelian gauge fields also track
the radiation during the radiation and matter eras.
Here, the difference from Ref.~\cite{gaugequ} is that 
such a tracking behavior ends at high redshifts ($z \gg 1$).

In all the cases shown in Fig.~\ref{fig3},
the solutions finally approach the isotropic de Sitter 
fixed point with a vanishing anisotropic hair. 
We have also run numerical simulations by choosing 
other model parameters and found that 
the property of decreasing $r_{\Sigma}$ and $v$
after the radiation domination is generic.
For $p={\cal O}(1)$ the approach of $w_{\rm DE}$ to 
$w_{\rm DE}^{({\rm iso})}$ occurs for 
$z \gg 1$, so the cosmological evolution 
at low redshifts is similar to that in the 
isotropic case.

\subsection{$G_{2,3,4,5} \neq 0, G_6 \neq 0, g_4=0, 
g_5 \neq 0, f_{4,5,6}=0$}
\label{secB}

If the terms $g_5$ and $G_6$ are present, they can modify 
the cosmological dynamics discussed in Sec.~\ref{secA}. 
For concreteness, we consider the following functions
\be
g_5(X)=\frac{2^{j_5-2}h_5}{mM_{\rm pl}^{2j_5+1}}X^{j_5}\,,
\qquad
G_6(X)=\frac{2^{p_6-1}h_6}{m^2M_{\rm pl}^{2p_6}}X^{p_6}\,,
\ee
where $h_5,j_5,h_6,p_6$ are dimensionless constants. 
We assume that both $j_5$ and $p_6$ are positive.

In the early cosmological epoch, the main contribution to the dark energy 
density originating from the spatial component $v$ corresponds 
to the term $-C_2H^2v^2$ in $C_4$. 
The terms $H\phi g_5$ and $H^2 G_6$, which appear in $C_2$ 
as well as in $q_V$, can be expressed as
\ba
H \phi g_5 &=&
h_5 \frac{\xi}{4} \left( \frac{\phi}{M_{\rm pl}} \right)^{2j_5+1-p}
\left( 1-\frac{v^2}{\phi^2} \right)^{j_5}\,,\\
H^2 G_6 &=& 
h_6 \frac{\xi^2}{2} \left( \frac{\phi}{M_{\rm pl}} 
\right)^{2p_6-2p} \left( 1-\frac{v^2}{\phi^2} \right)^{p_6}\,.
\ea
Provided that the parameter $\xi$ given by Eq.~(\ref{xi}) 
stays nearly constant 
around 1 and that $1-v^2/\phi^2$ is at most of the order of 1, 
the conditions that the terms $H\phi g_5$ and $H^2 G_6$ 
do not grow in the asymptotic past are given, respectively, by 
\be
j_5 \geq \frac12 (p-1)\,,\qquad p_6 \geq p\,.
\label{q5p6con}
\ee

The energy densities corresponding to the terms $4H\phi g_5$ 
and $-3H^2 G_6$ in $C_2$ are $\rho_{g_5}=-4\phi g_5v^2 H^3$ 
and $\rho_{G_6}=3G_6v^2 H^4$, respectively, so the associated 
density parameters $\Omega_{g_5}=\rho_{g_5}/(3M_{\rm pl}^2 H^2)$ 
and $\Omega_{G_6}=\rho_{G_6}/(3M_{\rm pl}^2 H^2)$ read
\be
\Omega_{g_5}=
-\frac{4H\phi g_5}{3} \frac{v^2}{M_{\rm pl}^2}\,,
\qquad
\Omega_{G_6}=
H^2 G_6 \frac{v^2}{M_{\rm pl}^2}\,.
\ee
Under the conditions (\ref{q5p6con}), we have that  
$|H\phi g_5| \ll 1$ and $|H^2 G_6| \ll 1$ 
in the early radiation era for the couplings $h_5, h_6$ 
at most of the order of unity.
In this case, both $|\Omega_{g_5}|$ and $|\Omega_{G_6}|$ 
are much smaller than 1 for $v^2 \lesssim M_{\rm pl}^2$. 
If the conditions (\ref{q5p6con}) are violated, the 
terms $(\phi/M_{\rm pl})^{2j_5+1-p}$ and 
$(\phi/M_{\rm pl})^{2p_6-2p}$ grow as we go 
back to the past. This leads to the values of 
$|\Omega_{g_5}|$ and $|\Omega_{G_6}|$ 
larger than 1 (unless we choose
very small values of $|h_5|$ and $|h_6|$).
To avoid this behavior, we need to impose 
the conditions (\ref{q5p6con}).

%%%%%%%%%%%%%%%%%%%%%%%%%%%%%%
\begin{figure}
\begin{center}
\includegraphics[height=3.4in,width=3.4in]{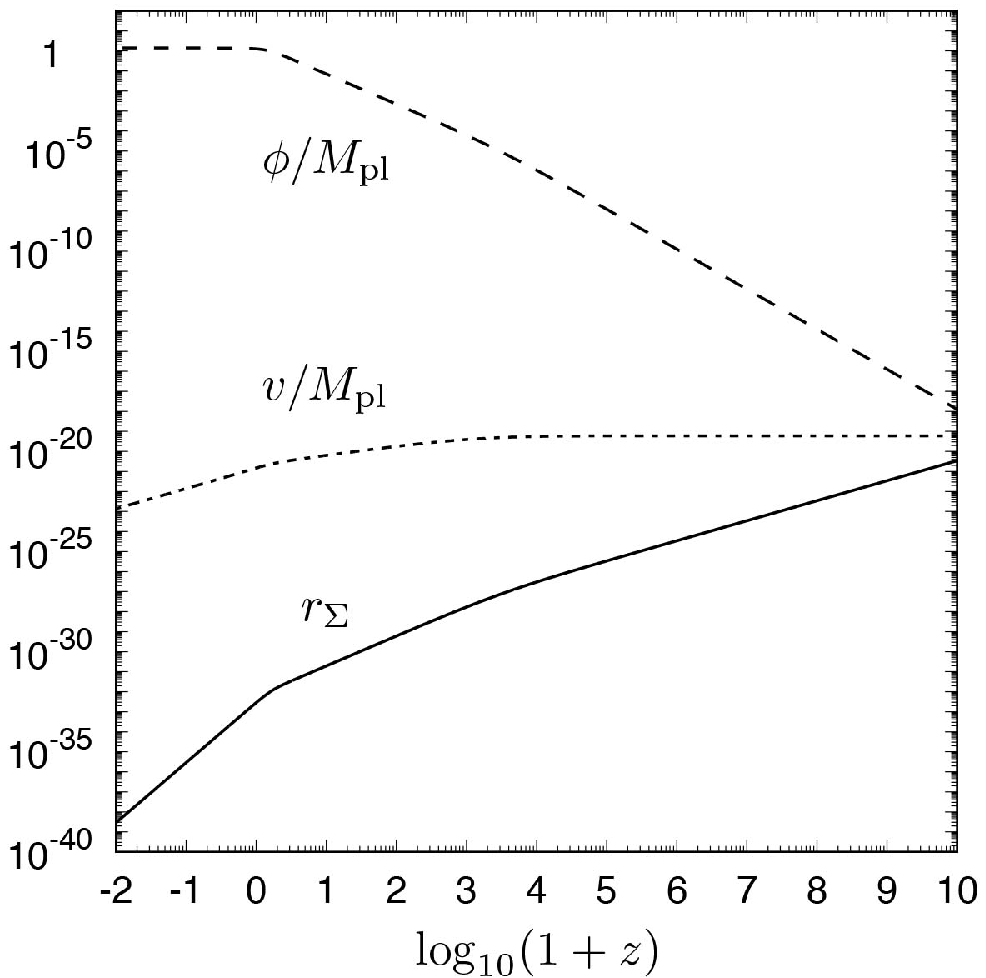}
\includegraphics[height=3.5in,width=3.4in]{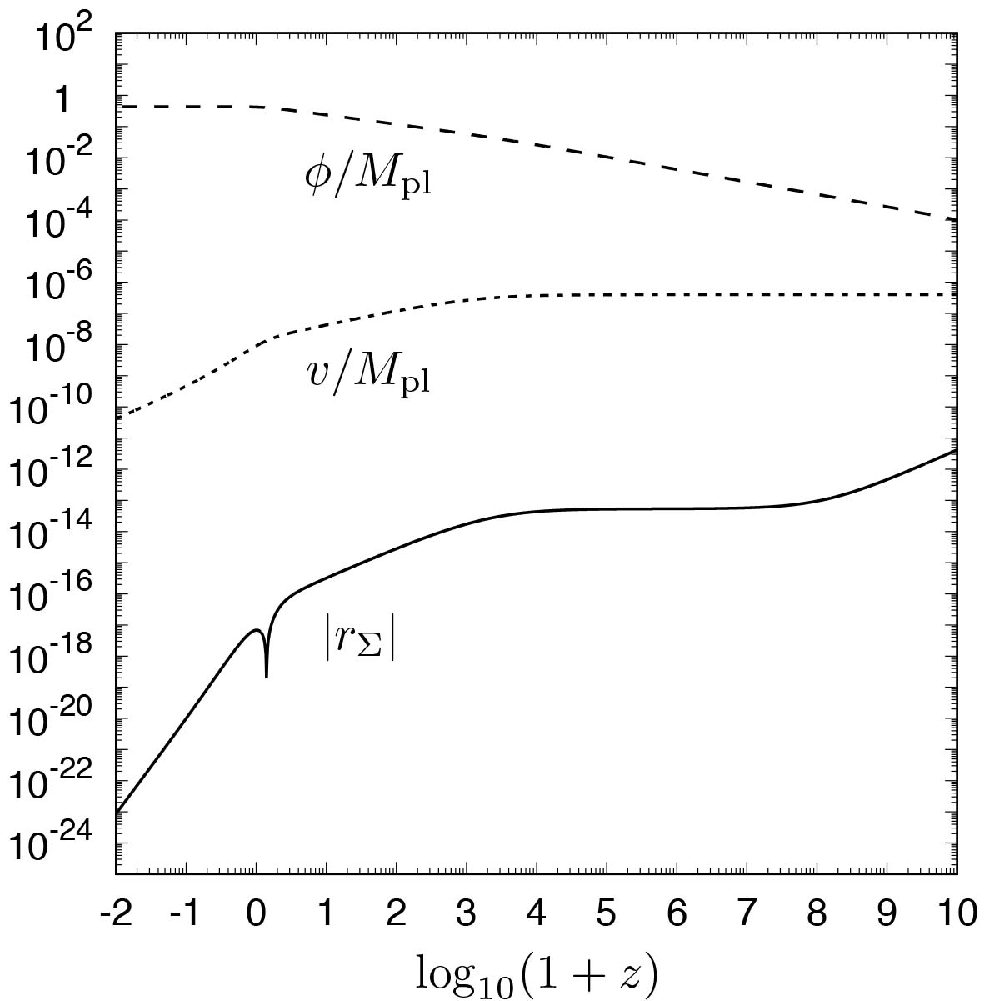}
\end{center}
\caption{\label{fig4}
Evolution of $\phi/M_{\rm pl}$, 
$v/M_{\rm pl}$ and $r_{\Sigma}$ in the model (B) for 
the parameters $p=1$, $a_4=0.01$, $a_5=0.05$, 
$j_5=1$, $h_5=-0.4$, $G_6=0$ (left) 
and $p=5$, $a_4=0.01$, $a_5=0.05$, 
$j_5=5$, $h_5=-4.0\times10^{-5}$, $G_6=0$ (right). 
The initial conditions are chosen to be
$\phi/M_{\rm pl}=6.0\times10^{-19}$, 
$v/M_{\rm pl}=6.0\times10^{-20}$, 
$\dot{v}=0$, $m\phi/(\sqrt{6} M_{\rm pl} H)=2.0\times10^{-37}$, 
$r_{\Sigma}=5.0\times10^{-21}$ and 
$1-\Omega_r=2.2\times10^{-7}$ 
at the redshift $z=1.4\times10^{10}$ (left), and  
$\phi/M_{\rm pl}=6.0\times10^{-5}$, 
$v/M_{\rm pl}=4.0\times10^{-7}$, 
$\dot{v}=0$, $m\phi/(\sqrt{6} M_{\rm pl} H)=2.0\times10^{-23}$, 
$r_{\Sigma}=1.0\times10^{-11}$ and 
$1-\Omega_r=1.3\times10^{-7}$ 
at $z=2.4\times10^{10}$ (right). 
}
\end{figure}
%%%%%%%%%%%%%%%%%%%%%%%%%%%%%%

Provided that $|H\phi g_5| \ll 1$ and $|H^2 G_6| \ll 1$, 
the quantity $q_V$ is close to 1 with $|\alpha_1/q_V| \ll 1$ 
in Eq.~(\ref{vapeq}).
Moreover the $g_5$ and $G_6$ dependent terms in $\alpha_2$ 
are suppressed relative to $q_V$, so the expression of $\alpha_2$ 
is similar to that given in Eq.~(\ref{alpha2}).
As long as $|\alpha_2/q_V| \ll 1$, the evolution of $v$ during 
the radiation era should be given by $v={\rm constant}$.
As for the anisotropic expansion rate during the radiation era, 
we have that $C_2 \simeq -1/2$ and $|H^2 C_5| \ll |HC_2|$ in 
Eq.~(\ref{Sigeq}) under the conditions $|H\phi g_5| \ll 1$ 
and $|H^2 G_6| \ll 1$, so we obtain the solution in the form 
(\ref{rsigso}) with $q_V \simeq 1$. 
This means that, unless $q_Vv^2/(3M_{\rm pl}^2)$ is much 
smaller than $|{\cal B}/(a^3H)|$ in the radiation-dominated epoch, 
$r_{\Sigma}$ temporally approaches the constant 
value $q_Vv^2/(3M_{\rm pl}^2)$.
If the condition $q_Vv^2/(3M_{\rm pl}^2) \ll |{\cal B}/(a^3H)|$ 
is always satisfied in the radiation era, then $r_{\Sigma}$ 
decreases as $r_{\Sigma} \propto a^{-1}$.

In Fig.~\ref{fig4} we plot two examples for the evolution of $\phi, v, r_{\Sigma}$ 
with the model parameters $h_5=-0.4$, $j_5=1$, $G_6=0$, and $p=1$ 
(left) and $h_5=-4.0\times10^{-5}$, $j_5=5$, $G_6=0$, and $p=5$ (right). 
The temporal vector component $\phi$ always increases 
during the radiation and matter eras.
As estimated analytically, the spatial vector component $v$ 
stays nearly constant in the radiation era and it decreases 
in proportion to $a^{-1/2}$ during the matter dominance. 
In the left panel of Fig.~\ref{fig4}, 
the initial value of $v$ is small such that 
the condition $q_Vv^2/(3M_{\rm pl}^2) \ll |{\cal B}/(a^3H)|$ 
is satisfied, so the ratio $r_{\Sigma}$ decreases as 
$\propto a^{-1}$ in the radiation era. 
In the right panel of Fig.~\ref{fig4}, the 
initial value of $v$ is larger than that for $p=1$, 
so $r_{\Sigma}$ temporally approaches the constant value 
$q_Vv^2/(3M_{\rm pl}^2)$. In both cases, $r_{\Sigma}$ 
decreases after the end of the radiation era.
In the numerical simulations of Fig.~\ref{fig4} 
the conditions (\ref{vin}) are initially satisfied with 
${\cal C}_{10} \simeq 2{\cal C}_2$, so the vector field 
temporally behaves as a dark radiation ($w_{\rm DE} \simeq 1/3$)
until $w_{\rm DE}$ approaches the isotropic value 
$w_{\rm DE}^{{\rm (iso)}}$.
As in the case of the model (A), the approach of $w_{\rm DE}$ 
to $w_{\rm DE}^{{\rm (iso)}}$ occurs at high redshifts ($z \gg 1$) 
for $p={\cal O}(1)$.

When $g_5<0$, the quantity ${\cal A}_V$ defined by Eq.~(\ref{AV}) 
is positive, so there exists the anisotropic de Sitter 
fixed point characterized by Eq.~(\ref{vc}).
As we showed analytically in Sec~\ref{deSittersec}, 
this anisotropic fixed point is not stable, while the 
isotropic de Sitter one is stable.
In fact, the numerical simulation of Fig.~4 (which corresponds to 
$g_5<0$) shows that the solution finally approaches the isotropic de Sitter 
fixed point characterized by $v=0$ and $\Sigma=0$, so the 
anisotropic hair does not survive. 

We also carry out numerical simulations for the model 
with non-zero $G_6$ and find that the cosmological evolution 
is qualitatively similar to that for $g_5 \neq 0$.
The general result is that, for cosmologically viable 
models with the late-time acceleration, the spatial vector 
component evolves as $v={\rm constant}$ (radiation era), 
$v \propto a^{-1/2}$ (matter era), 
$v \propto a^{-1}$ (de Sitter era) 
and that the ratio $r_{\Sigma}$ finally approaches 0.
The evolution of $w_{\rm DE}$ at low redshifts is similar 
to that in the isotropic case. 

\subsection{$f_{4,5,6} \neq 0$}
\label{bgsec}

Let us finally study the beyond-generalized Proca theories 
described by the action (\ref{action}) with ${\cal L}^{\rm N} \neq 0$.
Since $f_4 \neq 0$ and $f_5 \neq 0$ in such theories, 
the existence of coefficient $C_1$ in Eqs.~(\ref{eqN})-(\ref{eqp}) 
leads to the dynamical system (\ref{auto}) with different 
$Z$, ${\bm x}$, and ${\bm y}$.
In this case, after taking the time derivative of Eq.~(\ref{eqN}), 
we first eliminate the second derivative $\ddot{\phi}$ on account 
of Eq.~(\ref{eqa}). Similarly, the $\ddot{\phi}$ term is
eliminated by combining Eq.~(\ref{eqa}) with Eq.~(\ref{eqs}).
On using these two equations with Eq.~(\ref{eqv}), we can 
derive the two equations for $\dot{\Sigma}$ without containing 
$\ddot{v}$. Eliminating the $\dot{\Sigma}$ term, 
we obtain the following equation 
\be
v\,{\cal F}(\phi, \dot{\phi}, v, \dot{v}, H, \Sigma)=0\,,
\ee
where ${\cal F}$ is a function that depends on the quantities 
inside the parenthesis.
One of the branches of solutions is given by 
\be
v=0\,.
\label{v0}
\ee

Another branch corresponds to ${\cal F}=0$, which gives rise to 
a non-vanishing value of $v$.
Recall that the coefficient $C_1$ given in Eq.~(\ref{C1}) 
contains the term $v^2$.
If $v \neq 0$, then we can express $\dot{\phi}$ in terms 
of $\phi, v, \dot{v}, H, \Sigma$ by using Eq.~(\ref{eqN}).
Taking the time derivative of Eq.~(\ref{eqN}) and using 
Eqs.~(\ref{eqa}) and (\ref{eqs}), we obtain two equations 
after the elimination of $\ddot{\phi}$. 
On using other equations as well, the dynamical equations 
of motion can be written in the form (\ref{auto}), 
where ${\bm x}={}^t(\ddot{v},\dot{H},\dot{\Sigma})$, and
$Z, {\bm y}$ are the $3 \times 3$ and $1 \times 3$ matrices, 
respectively, involving the dependence of $\dot{v},v, \phi,H, \Sigma$. 
Unlike the theories with $f_{4,5}=0$, the $\dot{\phi}$ term 
is determined by the constraint equation (\ref{eqN}).
Computing the determinant of the matrix $Z$, 
we find that the determinant vanishes exactly. 
When the determinant vanishes we cannot solve Eq.~(\ref{auto}) 
in the form ${\bm x}=Z^{-1}{\bm y}$, 
so the dynamical system does not reduce to 
the closed autonomous system.
Note that this kind of determinant singularity 
also appears in the context of anisotropic string cosmology 
with dilaton and axion fields \cite{Topo}.

The above discussion shows that only the branch $v=0$
is physically allowed for the theories with $f_{4,5} \neq 0$. 
For this branch all the terms containing $C_1$ in 
Eqs.~(\ref{eqN})-(\ref{eqp}) vanish, so the dynamical 
system is similar to that of second-order 
generalized Proca theories with $v=0$. 
Then the anisotropic expansion rate simply decreases 
as $\Sigma \propto a^{-3}$.
If the ratio $|r_{\Sigma}|$ is much smaller than 1 at the onset 
of the radiation domination, the effect of $\Sigma$ on the dynamical 
equations of motion is negligible during most of 
the cosmic expansion history. 
In this sense, the cosmological dynamics 
for the theories with $f_{4,5} \neq 0$ is very similar to that 
of the isotropic case without having the $C_1$-dependent terms.

\section{Conclusions}
\label{consec}

In beyond-generalized Proca theories, we have studied the anisotropic 
cosmological dynamics in the presence of a spatial vector component $v$.
On the isotropic FLRW background it was found in Ref.~\cite{HKT} that, 
even with the Lagrangian density ${\cal L}^{\rm N}$ outside the domain 
of second-order generalized Proca theories, 
there is no additional DOF associated with the Ostrogradski ghost.
In this paper we showed that the same result also holds on the 
anisotropic background. 
There exists the constraint equation (\ref{Hamicon1}) related with the 
Hamiltonian ${\cal H}$ as Eq.~(\ref{LHN}), so that ${\cal H}=0$.
Hence the beyond-generalized Proca theories are 
free from the Ostrogradski instability with the 
Hamiltonian unbounded from below. 

In Sec.~\ref{anasecA} we analytically estimated the evolution of 
the anisotropic expansion rate $\Sigma$ and the spatial component 
$v$ in the early cosmological epoch.
If the conditions (\ref{Sigcon}) hold in the radiation and 
matter eras, $\Sigma$ decreases in proportion to $a^{-3}$.
Under the conditions (\ref{alcon}), the evolution of $v$ is 
given by $v={\rm constant}$ during the radiation era and 
$v \propto a^{-1/2}$ during the matter era.
If $v$ is not very much smaller than $M_{\rm pl}$ during the 
radiation domination, there are cases in which the second condition 
of Eq.~(\ref{Sigcon}) is violated. In concrete dark energy models studied 
in Sec.~\ref{numesec}, we showed the existence of solutions on 
which the ratio $r_{\Sigma}=\Sigma/H$ remains nearly constant 
during the radiation era. 

In Sec.~\ref{deSittersec} we discussed the property of de Sitter 
fixed points relevant to the late-time cosmic acceleration.
Besides the isotropic point (\ref{isofixed}), we found the 
existence of anisotropic fixed points (\ref{vc}) under the two 
conditions (\ref{dscon1}) and (\ref{dscon2}).
For the theories in which the parameter ${\cal A}_V$ defined by 
Eq.~(\ref{AV}) vanishes, we only have the isotropic fixed point.
For ${\cal A}_V \neq 0$ the anisotropic fixed points exist, 
but they are not stable. In both cases, the analytic estimation 
implies that the solutions approach the stable isotropic 
point in accordance with the cosmic no-hair conjecture.

In Sec.~\ref{numesec} we studied the evolution of anisotropic 
cosmological solutions in a class of dark energy models given by the 
functions (\ref{G2345}).
In the early cosmological epoch, the contributions of 
$v$ and $\Sigma$ to the energy density $\rho_{\rm DE}$ 
and the pressure $P_{\rm DE}$ can be larger than 
the isotropic contributions associated with the temporal 
vector component $\phi$.
If $v$ is large such that the conditions (\ref{vin}) are satisfied, 
the dark energy equation of state is given by Eq.~(\ref{wdees1}) 
in the radiation era, which is close to $w_{\rm DE}=1/3$
in concrete models studied in Secs.~\ref{secA} and \ref{secB}.
If the contribution of $\Sigma$ dominates over that of $v$ 
such that the conditions (\ref{vin2}) are satisfied,  
we have $w_{\rm DE} \simeq 1$ during the 
radiation era.
In both cases, the dark energy equation of state is 
different from the isotropic value 
$w_{\rm DE}^{(\rm iso)}$ given by Eq.~(\ref{wdeiso}).
However, the transition of $w_{\rm DE}$ to the 
value $w_{\rm DE}^{(\rm iso)}$ typically occurs 
at high redshifts (see Figs.~\ref{fig2} 
and \ref{fig3}), so the dark energy dynamics at low 
redshifts is similar to that in the isotropic case.

In generalized Proca theories with $v$ 
not very much smaller than $M_{\rm pl}$,
the spatial anisotropy in the radiation era can be sustained 
by $v$ with the nearly constant ratio 
$r_{\Sigma}\simeq q_Vv^2/(3M_{\rm pl}^2)$.
In this regime, for the models (A) and (B) studied in Sec.~\ref{numesec}, 
the vector field behaves as a dark radiation characterized 
by $w_{\rm DE} \simeq 1/3$.
As seen in the right panels of Fig.~\ref{fig1} and 
\ref{fig4}, the constant behavior of $r_{\Sigma}$ 
in the radiation era can occur for the models 
with large powers $p$ (like $p=5$) due to the possible
choice of large initial values of $v$. 
On the other hand, for the models with small $p$ (like $p=1$), 
we have $q_V v^2/(3M_{\rm pl}^2) \ll |{\cal B}/(a^3H)|$ 
in Eq.~(\ref{rsigso}) and hence $r_{\Sigma}$ decreases 
as $\propto a^{-1}$ during the radiation era 
(see the left panels of Fig.~\ref{fig1} and \ref{fig4}). 
After the matter dominance, both $v$ and 
$\Sigma$ decrease toward the isotropic fixed point 
($v=0=\Sigma$).

In beyond-generalized Proca theories, we showed that the 
physical branch of solutions without having a determinant 
singularity of the dynamical system corresponds to $v=0$. 
In this case the anisotropic expansion rate simplify 
decreases as $\Sigma \propto a^{-3}$ from the onset of 
the radiation-dominated epoch, so the cosmological evolution 
is practically indistinguishable from the isotropic case.
Interestingly, the beyond-generalized Proca theories do not 
allow the existence of anisotropic solutions with 
constant $r_{\Sigma}$.

We have thus shown that, apart from the radiation era 
in the presence of a non-negligible spatial vector component $v$, 
the anisotropy does not survive for a class of dark 
energy models in the framework of (beyond-)generalized Proca theories. 
Thus, the analysis of Refs.~\cite{cosmo,Geff} where
the spatial component was treated as a perturbation 
on the isotropic FLRW background can be justified except for 
the early cosmological epoch in which the vector field 
behaves as a dark radiation. 
It will be of interest to place detailed observational constraints on both 
isotropic and anisotropic dark energy models from the observations of 
CMB, type Ia supernovae, and large-scale structures.

\section*{Acknowledgements}

LH thanks financial support from Dr.~Max R\"ossler, 
the Walter Haefner Foundation and the ETH Zurich
Foundation. RK is supported by the Grant-in-Aid for Research Activity
Start-up of the JSPS No.\,15H06635. 
ST is supported by the Grant-in-Aid for Scientific Research Fund of 
the JSPS No.\,16K05359 and MEXT KAKENHI Grant-in-Aid for 
Scientific Research on Innovative Areas ``Cosmic Acceleration'' 
(No.\,15H05890).

%%%%%%%%%
\appendix
%%%%%%%%%

%
\section{Coefficients of equations of motion}

The coefficients appearing in Eqs.~(\ref{eqa})-(\ref{eqv})
are given by 
\ba
&&
C_{5}=4 \phi g_{5}-4 ( H+\Sigma ) \left[ G_{6}+{\phi}^{2} ( G_{6,X}+2 f_{6} ) \right] 
\,,\notag\\
&&
C_{6}=2 v \left[ C_{2}-6 \phi \Sigma g_{5}-2 G_{4,X}-2 \phi^{2}f_{4}
- ( H+\Sigma ) \left\{\phi ( 6 \phi^{2}f_{5}-G_{5,X} ) 
-6 \Sigma ( G_{6} +\phi^{2}G_{6,X}+2 \phi^{2}f_{6})\right\} \right]
\,,\notag\\
&&
C_{7}=12 \phi^{2}v^{2}f_{5}
\,,\quad
C_{8}=-2 \left[G_{6}+ \phi^{2} ( G_{6,X}+2 f_{6} )\right] 
\,,\notag\\
&&
C_{9}=2 v \left[ \phi ( 4 g_{5}+G_{5,X}-6 \phi^{2}f
_{5} ) -6H ( G_{6}+\phi^{2}G_{6,X}+2 \phi^{2}f_{6} ) \right]
\,,\notag\\
&&
C_{10}=12 (G_{4}-\phi^{2} G_{4,X}- \phi^{4}f_{4})
+6 \phi^{3}H ( G_{5,X}-6 \phi^{2}f_{5} ) 
\notag\\
&&\hspace{1cm}
-v^{2} \left[ 1-2 g_{4}-12 \phi^{2}f_{4}-36 H\phi^{3}f_{5}-12 \phi ( H-\Sigma ) g_{5}  
+6 ( 2 {H}^{2}-2 H\Sigma-{\Sigma}^{2} ) ( G_{6}+\phi^{2}G_{6,X}+2 \phi^{2}f_{6} ) \right]
\,,\notag\\
&&
C_{11}=2 v \left[ \phi ( G_{5,X}-2 g_{5}-6 \phi^{2}f_{5} ) 
+6 \Sigma ( G_{6}+\phi^{2}G_{6,X}+2 \phi^{2}f_{6} ) \right]
\,,\notag\\
&&
C_{12}=6 \phi^{3}\Sigma ( 6 \phi^{2}f_{5}-G_{5,X} ) 
+2 v^{2} \left[ 1-2 g_{4}-18 \phi^{3}\Sigma f_{5}-6 \phi H g_{5}
 +3 ( {H}^{2}+2 H\Sigma-2 {\Sigma}^{2} ) ( G_{6}+\phi^{2}G_{6,X}+2\phi^{2} f_{6})   \right] 
\,,\notag\\
&&
C_{13}= 4v^{2}\left[ f_{4}+\phi^{2}f_{4,X}
+3 \phi ( H+\Sigma ) ( 2 f_{5}+\phi^{2}f_{5,X} ) \right] 
\,,\notag\\
&&
C_{14}=2 g_{5}+2 \phi^{2}g_{5,X}-2 \phi ( H+\Sigma ) 
\left[ 3 G_{6,X}+4 f_{6}+\phi^{2}( G_{6,{XX}}+2 f_{6,X} )\right]  
\,,\notag\\
&&
C_{15}=2 v \big[ \phi g_{4,X}+2 ( 2 H-\Sigma ) ( g_{5}+\phi^{2}g_{5,X} ) 
-2 \phi ( G_{4,{XX}}+\phi^{2}f_{4,X} ) 
\notag\\
&&\hspace{1cm}
+( H+\Sigma ) \left\{ G_{5,X}+\phi^{2}G_{5,{XX}}-6 \phi^{2} (f_{5}+\phi^{2}f_{5,X}) \right\} 
\notag\\
&&\hspace{1cm}
-3 \phi ( H^2-\Sigma^2 ) ( 3 G_{6,X}+\phi^{2}G_{6,{XX}}+4 f_{6}+2 \phi^{2}f_{6
,X} ) -2 \phi v^{2} \left\{ f_{4,X}+3 \phi ( H+\Sigma ) f_{5,X} \right\}  \big]
\,,\notag\\
&&
C_{16}=-3 \phi^{2} G_{3,X}-12 \phi H \left[ G_{4,X}+\phi^{2}G_{4,{XX}}
+\phi^{2} (4f_{4}+\phi^{2}f_{4,X}) \right] 
\notag\\
&&\hspace{1cm}
+3 \phi^{2} ( H^2-\Sigma^2 ) \left[ 3 G_{5,X}+\phi^{2}G_{5,{XX}}-6 \phi^{2} (5f_{5}+ {\phi}^{2}f_{5,X}) \right]
\notag\\
&&\hspace{1cm}
+v^{2} \big[ 2 ( H-2 \Sigma ) \left\{ \phi g_{4,X}+3 H ( g_{5}+\phi^{2}g_{5,X} ) \right\}  
+12 \phi \left\{ ( 2 H-\Sigma ) f_{4}
+H\phi^{2}f_{4,X} \right\} 
\notag\\
&&\hspace{1cm}
+18 \phi^{2} ( H^2-\Sigma^2 ) ( 4 f_{5}+\phi^{2} f_{5,X} ) 
-2 \phi ( 2 H-\Sigma ) ( H-2 \Sigma ) ( H+\Sigma ) ( 3 G_{6,X}+\phi^{2}G_{6,{XX}}+4 f_{6}+2\phi^{2}f_{6,X} ) \big] 
\,,\notag\\
&&
C_{17}=-2 v \left[ \phi g_{5,X}-( H+\Sigma ) \left\{ G_{6,X}+\phi^{2} (G_{6,{XX}}+2 f_{6,X}) \right\}  \right]
\,,\notag\\
&&
C_{18}=2 v^{2} \left[ 2 G_{4,{XX}}-g_{4,X}+2 \phi^{2}f_{4,X}
+\phi ( H+\Sigma ) ( 6 \phi^{2}f_{5,X}-G_{5,{XX}} ) 
- ( H+\Sigma ) ^{2} ( G_{6,X}+\phi^{2}G_{6,{XX}}+2 \phi^{2}f_{6,X} )  \right] 
\notag\\
&&\hspace{1cm}
+2 v ( 2 H-\Sigma ) C_{17}-C_{2}
+\frac32 ( 2 H-\Sigma ) C_{5}
-4 G_{4,X}-4 \phi^{2}f_{4}-2 \phi ( H+\Sigma) ( 6 \phi^{2}f_{5}-G_{5,X} )  
\,,\notag\\
&&
C_{19}=
-2 v^3\left[ 6 H\phi^{2}f_{4,X}+9 \phi^{3} ( {H}^{2}-{\Sigma}^{2} ) f_{5,X} 
+( H -2 \Sigma ) \left\{g_{4,X} + ( H+\Sigma )^{2} ( G_{6,X} +\phi^{2}G_{6,{XX}}+2 \phi^{2}f_{6,X})\right\} \right]
\notag\\
&&\hspace{1cm}
+3 v^2 H ( H -2 \Sigma ) C_{17}+v \Big[ 3 \phi G_{3,X}
+4 ( H+\Sigma ) C_{2}+2 ( {H}^{2}-7 H\Sigma+{\Sigma}^{2} ) C_{5}
-12 ( H-\Sigma ) G_{4,X}
\notag\\
&&\hspace{1cm}
+12 \phi^{2} ( 2 H+\Sigma ) f_{4}+12 \phi^{2}H ( \phi^{2}f_{4,X}+G_{4,{XX}} ) 
+3 \phi  ( {H}^{2}-{\Sigma}^{2} ) \left\{ G_{5,X}-{\phi}^{2} (G_{5,{XX}}-6 f_{5}-6 \phi^{2}f_{5,X}) \right\} 
\Big] 
\,,\notag\\
&&
C_{20}=3 G_{2}+18 ( {H}^{2}+{\Sigma}^{2} ) ( G_{4}-\phi^{2}G_{4,X}-{\phi}^{4}f_{4} ) 
-6 \phi^{3} ( {H}^{3}+{\Sigma}^{3} ) ( 6 \phi^{2}f_{5}-G_{5,X} ) 
\notag\\
&&\hspace{1cm}
+ v^2 \left[ 3 ( {H}^{2}-4 {\Sigma}^{2} ) C_{2}-\frac32 \Sigma ( 5 H+2 \Sigma ) ( H -2 \Sigma) C_{5}
+18 \phi^{2} ( {H}^{2}+{\Sigma}^{2} ) f_{4}+36 \phi^{3} ( {H}^{3}+{\Sigma}^{3} ) f_{5}
\right]
\,,\notag\\
&&
C_{21}= \left[ 3 ( H -2 \Sigma ) C_{5}-12 G_{4,X}-12 \phi^{2}f_{4}
-6 \phi ( H+\Sigma) ( 6 \phi^{2}f_{5}-G_{5,X} )  \right] v-2 C_{6}
\,,\notag\\
&&
C_{22}= v^2 \left[ 4 ( 2 \Sigma-H ) C_{2}+\frac12 ( 5 H+2 \Sigma ) ( H -2 \Sigma ) C_{5}
-12 \phi^{2}\Sigma \left\{ f_{4}+3 \phi ( H+\Sigma ) f_{5}  \right\} \right]
 \notag\\
&&\hspace{1cm}
 -6 \Sigma \left[ 2 G_{4}-2 \phi^{2}G_{4,X}-2 \phi^{4}f_{4}+\phi^{3} ( H+\Sigma ) 
 ( G_{5,X}-6 \phi^{2}f_{5} ) \right] \,, \notag\\
&& 
D_{1}=-\phi g_{4,X}-2 ( H+\Sigma ) ( g_{5}+\phi^{2}g_{5,X} ) 
+\phi ( H+\Sigma )^{2} ( 3 G_{6,X}+\phi^{2}G_{6,{XX}}+4 f_{6}+2 \phi^{2}f_{6,X} ) 
\,,\notag\\
&&
D_{2}= v \big[ G_{3,X}-2 \phi ( H-2\Sigma ) g_{4,X}  + 
( H+\Sigma ) \big\{ 4 \phi ( G_{4,{XX}}+4 f_{4}+\phi^{2}f_{4,X} ) 
-4 ( H-2\Sigma) ( g_{5}+\phi^{2}g_{5,X} ) \big\} 
 \notag\\
&&\hspace{1cm}
- ( H+\Sigma )^2\big\{  G_{5,X}+\phi^{2}G_{5,{XX}}-30 \phi^{2}f_{5}-6 \phi^{4}f_{5,X}  
-2 \phi ( H-2\Sigma) ( 3 G_{6,X}+\phi^{2}G_{6,{XX}}+4 f_{6}+2\phi^{2}f_{6,X} ) \big\} 
 \big] 
 \notag\\
&&\hspace{1cm}
 -2 \phi v^{3} ( H+\Sigma ) \left[ 2 f_{4,X}+3 \phi ( H+\Sigma ) f_{5,X} \right]
\,,\notag\\
&&
D_{3}=\phi ( G_{2,X}+3 H\phi G_{3,X} ) 
+6 \phi ( H^2-\Sigma^2 ) ( G_{4,X}+\phi^{2}G_{4,{XX}}+4 \phi^{2}f_{4}+\phi^{4}f_{4,X} ) 
 \notag\\
&&\hspace{1cm}
-\phi^{2} ( H+\Sigma )^{2} ( H-2\Sigma ) ( 3 G_{5,X}+\phi^{2}G_{5,{XX}}-30 \phi^{2}f_{5}-6 \phi^{4}f_{5,X} ) 
 \notag\\
&&\hspace{1cm}
-v^{2} \big[ ( H-2\Sigma ) ^{2} \left\{ \phi g_{4,X}+2 ( H+\Sigma ) ( g_{5}+\phi^{2}g_{5,X}) \right\}
-6 \phi ( H+\Sigma ) \left\{2 \Sigma f_{4} -(H-\Sigma)\phi^{2}f_{4,X} \right\} 
 \notag\\
&&\hspace{1cm}
-( H+\Sigma )^{2} \left\{6 \phi^{2} (6 \Sigma f_{5}- \phi^{2}(H-2\Sigma) f_{5,X} ) 
+\phi ( H-2 \Sigma ) ^{2} ( 3 G_{6,X}+\phi^{2}G_{6,{XX}}+4 f_{6} +2 \phi^{2}f_{6,X})\right\} \big] 
\,,\notag\\
&&
D_{4}=1-2 g_{4}-4 ( H+\Sigma) \phi g_{5} 
+2 ( H+\Sigma ) ^{2} (G_{6}+\phi^{2}G_{6,X}+2 \phi^{2}f_{6} ) 
\,,\notag\\
&&
D_{5}= v\left[g_{4,X} +2 \phi ( H+\Sigma ) g_{5,X}
- ( H+\Sigma ) ^{2} ( G_{6,X} +\phi^{2}G_{6,{XX}}+2 \phi^{2}f_{6,X}) \right] \,, \notag
\ea
\ba
\hspace{-1cm}&&
D_{6}= v \big[ G_{2,X}+3 \phi H G_{3,X}
+2 ( H+\Sigma ) ( H-2\Sigma ) \left\{ 1-2 g_{4}-4 \phi ( H+\Sigma ) g_{5} \right\} 
 \notag\\
&&\hspace{1cm}
+6 ( H+\Sigma )  
\left\{ ( H+\Sigma ) G_{4,X}+2 \phi^{2} ( 2 H-\Sigma ) f_{4}
+\phi^{2} ( H-\Sigma ) ( G_{4,{XX}}+{\phi}^{2}f_{4,X} ) \right\} 
 \notag\\
&&\hspace{1cm}
-\phi ( H+\Sigma )^{2} \left\{ 3 HG_{5,X}-6 \phi^{2} ( 5 H-4\Sigma ) f_{5}
+\phi^{2} ( H-2\Sigma ) ( G_{5,{XX}}-6 \phi^{2} f_{5,X} ) \right\}
 \notag\\
&&\hspace{1cm}
+4 ( H+\Sigma ) ^{3} ( H-2\Sigma ) ( G_{6}+\phi^{2}G_{6,X}+2 {\phi}^{2}f_{6} ) \big] 
- {v}^{3}\big[ ( H-2\Sigma )^{2} ( g_{4,X}+2 \phi ( H+\Sigma ) g_{5,X} ) 
 \notag\\
&&\hspace{1cm}
+6 \phi^{2} ( H^2-\Sigma^2 ) f_{4,X} 
+( H+\Sigma )^{2} ( H-2\Sigma )\left\{ 6 \phi^{3}  f_{5,X} 
- ( H-2\Sigma ) ( G_{6,X} +\phi^{2}G_{6,{XX}}+2 \phi^{2}f_{6,X})\right\} \big] \,.
\ea

\end{document}